\documentclass[letterpaper,times]{IONconf}

\usepackage[justification=centering]{caption}
\usepackage[noadjust]{cite}
\usepackage{hyperref}
\usepackage{amsmath,amssymb,amsfonts}
\usepackage[ruled,vlined]{algorithm2e}
\usepackage{graphicx}
\usepackage{textcomp}
\usepackage[binary-units]{siunitx}
\usepackage{lscape}
\usepackage[table]{xcolor}
\usepackage{tabularx}
\usepackage{rotating}
\usepackage{multicol}
\usepackage{subcaption}
\usepackage{bm}
\usepackage{float}
\usepackage{multirow}

\usepackage{url}

\usepackage{glossaries}

\newacronym{dhcp}{DHCP}{Dynamic Host Configuration Protocol}
\newacronym{ntp}{NTP}{Network Time Protocol}
\newacronym{tcp}{TCP}{Transmission Control Protocol}
\newacronym{udp}{UDP}{User Datagram Protocol}
\newacronym{crc}{CRC}{Cyclic Redundancy Check}
\newacronym{cdn}{CDN}{Content Delivery Network}
\newacronym{vpn}{VPN}{Virtual Private Network}

\newacronym{rss}{RSS}{Received Signal Strength}
\newacronym{mimo}{MIMO}{Multiple Input Multiple Output}
\newacronym{ofdm}{OFDM}{Orthogonal Frequency Division Multiplexing}
\newacronym{dsss}{DSSS}{Direct Sequence Spread Spectrum}
\newacronym{css}{CSS}{Chirp Spread Spectrum}
\newacronym{fsk}{FSK}{Frequency Shift Keying}
\newacronym{snr}{SNR}{Signal-to-Noise Ratio}
\newacronym{sinr}{SINR}{Signal to interference plus noise ratio}
\newacronym{aoa}{AoA}{Angle of Arrival}
\newacronym{ssc}{SSC}{Secret Spreading Code}
\newacronym{ci}{CI}{Constructive Interference}
\newacronym{cfo}{CFO}{Carrier Frequency Offset}
\newacronym{cpo}{CPO}{Carrier Phase Offset}
\newacronym{sf}{SF}{Spreading Factor}
\newacronym{cdma}{CDMA}{Code-division multiple access}
\newacronym{fdma}{FDMA}{Frequency division multiple access}
\newacronym{prn}{PRN}{Pseudorandom Noise}
\newacronym{bpsk}{BPSK}{Binary Phase-Shift Keying}
\newacronym{qam}{QAM}{Quadrature Amplitude Modulation}
\newacronym{boc}{BOC}{Binary Offset Carrier}
\newacronym{mboc}{MBOC}{Multiplexed Binary Offset Carrier}
\newacronym{tmboc}{TMBOC}{Time Multiplexed Binary Offset Carrier}
\newacronym{cboc}{CBOC}{Composite Binary Offset Carrier}

\newcommand\rar{Record-and-Replay}
\newcommand\gpssdrsim{GPS-SDR-SIM}
\newcommand\gnsssdr{GNSS-SDR}
\newcommand\gnuradio{GNU Radio}

\newacronym{arx}{ARX}{Adversarial Receiver}
\newacronym{atx}{ATX}{Adversarial Transmitter}
\newacronym{rra}{RRA}{Record-and-Replay-Attack}
\newacronym{dd}{DD}{Distance-Decreasing}
\newacronym{aua}{AuA}{Area under Attack}
\newacronym{dos}{DoS}{Denial of Service}
\newacronym{pki}{PKI}{Public Key Infrastructure}
\newacronym{tesla}{TESLA}{Timed Efficient Stream Loss-Tolerant Authentication}

\newacronym{gnss}{GNSS}{Global Navigation Satellite System}
\newacronym{gps}{GPS}{Global Positioning System}
\newacronym{nmea}{NMEA}{National Marine Electronics Association}
\newacronym{nasa}{NASA}{National Aeronautics and Space Administration}
\newacronym{agps}{A-GPS}{Assisted/Augmented GPS}
\newacronym{scer}{SCER}{Security Code Estimation and Replay}
\newacronym{nma}{NMA}{Navigation Message Authentication}
\newacronym{os-nma}{OS-NMA}{Open Service Navigation Message Authentication}
\newacronym{sce}{SCE}{Spreading Code Encryption}
\newacronym{sis}{SIS}{Signal In Space}
\newacronym{raim}{RAIM}{Receiver autonomous integrity monitoring}
\newacronym{pvt}{PVT}{Position-Velocity-Time}
\newacronym{pnt}{PNT}{Positioning, Navigation and Timing}
\newacronym{dop}{DoP}{Dilution of Precision}
\newacronym{glonass}{GLONASS}{Globalnaja nawigazionnaja sputnikowaja sistema}
\newacronym{qzss}{QZSS}{Quasi-Zenith Satellite System}
\newacronym{bds}{BDS}{BeiDou Navigation Satellite System}
\newacronym{irnss}{IRNSS}{Indian Regional Navigation Satellite System}
\newacronym{navic}{NavIC}{Navigation with Indian Constellation}
\newacronym{dgps}{DGPS}{Differential Global Positioning System}
\newacronym{ttff}{TTFF}{Time To First Fix}
\newacronym{tsa}{TSA}{Time Synchronization Attack}
\newacronym{ecef}{ECEF}{Earth-centered, Earth-fixed}
\newacronym{sca}{SCA}{Spreading Code Authentication}
\newacronym{tow}{TOW}{Time of Week}

\newacronym{sdr}{SDR}{Software Defined Radio}
\newacronym{cots}{COTS}{Commercially off the Shelf}
\newacronym{rt}{RT}{Real-Time}
\newacronym{ide}{IDE}{Integrated Development Environment}
\newacronym{fifo}{FIFO}{First-In-First-Out}
\newacronym{sma}{SMA}{SubMiniature Version A}
\newacronym{rf}{RF}{Radio Frequency}
\newacronym{imu}{IMU}{Inertial Measurement Unit}
\newacronym{usb}{USB}{Universal Serial Bus}
\newacronym{pll}{PLL}{Phase-Locked Loop}
\newacronym{lna}{LNA}{Low-Noise Amplifier}
\newacronym{tcxo}{TCXO}{Temperature Controller Oscillator}

\newacronym{foi}{FoI}{Feature of Interest}

\usepackage[capitalise, noabbrev]{cleveref}

\title{Distributed and Mobile Message Level Relaying/Replaying of GNSS Signals}

\author{
   Malte Lenhart, Marco Spanghero, Panos Papadimitratos \\%
   \textit{Networked Systems Security Group, KTH Royal Institute of Technology, Sweden}
}

\begin{document}

\maketitle

\section*{biography}
\biography{Malte Lenhart}{received his B.Sc. in Information System Technologies from Technical University of Darmstadt, Germany, where he is currently pursuing his M.Sc. focused on IT security. He is currently writing his master thesis on GNSS security at the Networked Systems Security Group at the KTH Royal Institute of Technology, Stockholm, Sweden, under the supervision of Marco Spanghero and Panos Papadimitratos.}

\biography{Marco Spanghero}{received his B.Sc. in Electronics Engineering from Politecnico of Milano and M.Sc. degree in Embedded System from KTH Royal Institute of Technology, Stockholm, Sweden. He is currently a Ph.D. candidate with the Networked Systems Security (NSS) group at KTH Royal Institute of Technology, Stockholm, Sweden. His research is concerned with secure positioning and synchronization.}

\biography{Panos Papadimitratos}{is a professor with the School of Electrical Engineering and Computer Science (EECS) at KTH Royal Institute of Technology, Stockholm, Sweden, where he leads the Networked Systems Security (NSS) group. He earned his Ph.D. degree from Cornell University, Ithaca, New York, in 2005. His research agenda includes a gamut of security and privacy problems, with emphasis on wireless networks. He is an IEEE Fellow, an ACM Distinguished Member, and a Fellow of the Young Academy of Europe.
}

\section*{Abstract}
With the introduction of \gls{nma}, future \glspl{gnss} prevent spoofing by simulation, i.e., the generation of forged satellite signals based on publicly known information. However, authentication does not prevent record-and-replay attacks, commonly termed as meaconing. Meaconing attacks are less powerful in terms of adversarial control over the victim receiver location and time, but by acting at the signal level, they are not thwarted by \gls{nma}. This makes replaying/relaying attacks a significant threat for current and future GNSS. While there are numerous investigations on meaconing attacks, the vast majority does not rely on actual implementation and experimental evaluation in real-world settings. 
In this work, we contribute to the improvement of the experimental understanding of meaconing attacks. We design and implement a system capable of real-time, distributed, and mobile meaconing, built with off-the-shelf hardware. We extend from basic distributed meaconing attacks, with signals from different locations relayed over the Internet and replayed within range of the victim receiver(s). 
This basic attack form has high bandwidth requirements and thus depends on the quality of service of the available network to work. 
To overcome this limitation, we propose to replay on message level, i.e., to demodulate and re-generate signals before and after the transmission respectively (including the authentication part of the payload). The resultant reduced bandwidth enables the attacker to operate in mobile scenarios, as well as to replay signals from multiple GNSS constellations and/or bands simultaneously. Additionally, the attacker can delay individually selected satellite signals to potentially influence the victim position and time solution in a more fine-grained manner. 
Our versatile test-bench, enabling different types of replaying/relaying attacks, facilitates testing realistic scenarios towards new and improved replaying/relaying-focused countermeasures in GNSS receivers. 

\section{Introduction}
\gls{gnss} receivers are extensively used to provide precise location and time to a wide gamut of applications, including critical infrastructures. Due to the intrinsic lack of security in the \gls{gnss} civilian user segment, adversarial modification of \gls{gnss}-provided \gls{pvt} solutions is a concrete risk \cite{gnss-jamming-blacksea, gnss-jamming-hype-article}. While military-grade \gls{gnss} signals are protected (to a certain extent) from spoofing, leveraging encrypted spreading codes, the structure and content of civilian signals, which are within the scope of this work, are publicly known and unprotected. 
Attackers can therefore spoof signals by simulating and transmitting signals that appear to originate from legitimate satellites. Simulation spoofing provides fine-grained control over the \gls{pvt} solutions at the victim receiver that is misled to lock on them. Open-source software to simulate \gls{gps} signals is publicly available \cite{ebinumaOsqzssGpssdrsim2020}, and an earlier work on an intermediate \gls{gps} spoofer proved to be effective against civilian receivers, yielding a great level of control over the victim's device with low detection probability \cite{humphreys2008assessing}.

The introduction of authentication codes for civilian messages through Galileo \gls{os-nma} \cite{Fernandez-Hernandez2016} or \gls{sca} in \gls{gps} \cite{Anderson2017}, scheduled for testing in December 2021 and 2022 respectively, will prevent attacks based on signal simulation: attackers would lack the necessary cryptographic keys for the authentication payload. This raises the bar for attackers, who still are able to resort to relay/replay (meaconing) attacks: recording signals in one location and replaying them back at a later time and, commonly, at a different location\footnote{Replaying can be used for legitimate purposes, for instance, to provide \gls{gnss} coverage inside aircraft hangers, by repeating roof-top antenna signals. However, even such legitimate use cases can turn into unintentional meaconing incidents, if, for instance, the hangar door is left open and the replayed signal provides \gls{gnss} receivers outside the hangar with conflicting location information \cite{gnss-jamming-newark, Coulon2020}.}.
Replaying the \gls{gnss} spectrum can induce an attacker-chosen position and time to the victim receiver by means of jamming and replaying with a power advantage \cite{Papadimitratos2008c}.
\gls{scer} attacks show the capabilities of replaying with respect to tampering with authenticated signals and constitute one the most advanced type of spoofing attacks \cite{Humphreys2013}.

In this work we focus on implementing practical replay/relay attacks at the signal and message level with colluding and distributed attacker nodes. Extending the basic (distributed) meaconing attack that replays an entire radio spectrum band, we show that replaying at message level reduces the required bandwidth. 
Our proposed approach demodulates legitimate signals and re-generates replicas used for spoofing. Our system is designed to replay/relay authenticated signals with small adaptations once they become publicly observable.
Our modular test-bench is capable to launch one-to-many distributed replay attacks, enabling us to test attack effects on mobile victims using multiple network-connected colluding adversarial nodes.

The remainder of this paper is structured as follows: \cref{sec:related-work} discusses relevant related work and \cref{sec:attacker-model} presents the adversary model considered in this work. \cref{sec:implementation,sec:experimental} present the proposed implementation for signal and message level replaying/relaying system, and the experimental setup used to demonstrate its effectiveness. \cref{sec:evaluation} evaluates our proposed design. \cref{sec:conclusion}  concludes this work with a discussion of possible future developments.

\section{Related Work}
\label{sec:related-work}

\gls{gnss} receivers calculate the \gls{pvt} solution based on visible  satellites at a point in time. Due to transmitted power and attenuation over long distances between the satellite orbits and the Earth surface, the received power of the \gls{gnss} signals is in the order of \SI{-180}{\decibel m}. For this reason, \gls{gnss} signals are susceptible to interference, adversarial overpowering, and environmental factors, such as multipath effects, shielding, and weather. Moreover, simple detection schemes based on received signal power are of limited use, in terms of attack detection reliability and confidence \cite{schmidtSurveyAnalysisGNSS2016}. To guarantee best service availability, \gls{gnss} receivers prioritize robustness and accuracy: this causes the receiver to lock onto the \gls{gnss} signals that show the highest signal-to-noise ratio, without considering their provenance. From an attacker perspective, this is favorable: once the receiver starts tracking a satellite, transmitting at a power advantage prevents the victim to switch back to tracking legitimate signals. 

As the majority of commercial \gls{gnss} receivers lack fundamental security features, \gls{gnss} receivers are often victims of various attacks, notably jamming and spoofing. While jamming is well documented and more straightforward to detect, spoofing is harder to detect and attribute. This is especially true for advanced spoofing attacks, such as the \gls{tsa}, which lifts victims off legitimate satellite signals without relying on jamming to cause a loss of lock \cite{humphreys2008assessing}. Without the sophistication of such advanced attacks, combined jamming and spoofing are likely to 'capture' commercial \gls{gnss} receivers, undetected, without triggering any alarm. 

\begin{figure*}[htb]
  \centering
  \begin{subfigure}[t]{0.44\textwidth}
    \includegraphics[width=\linewidth]{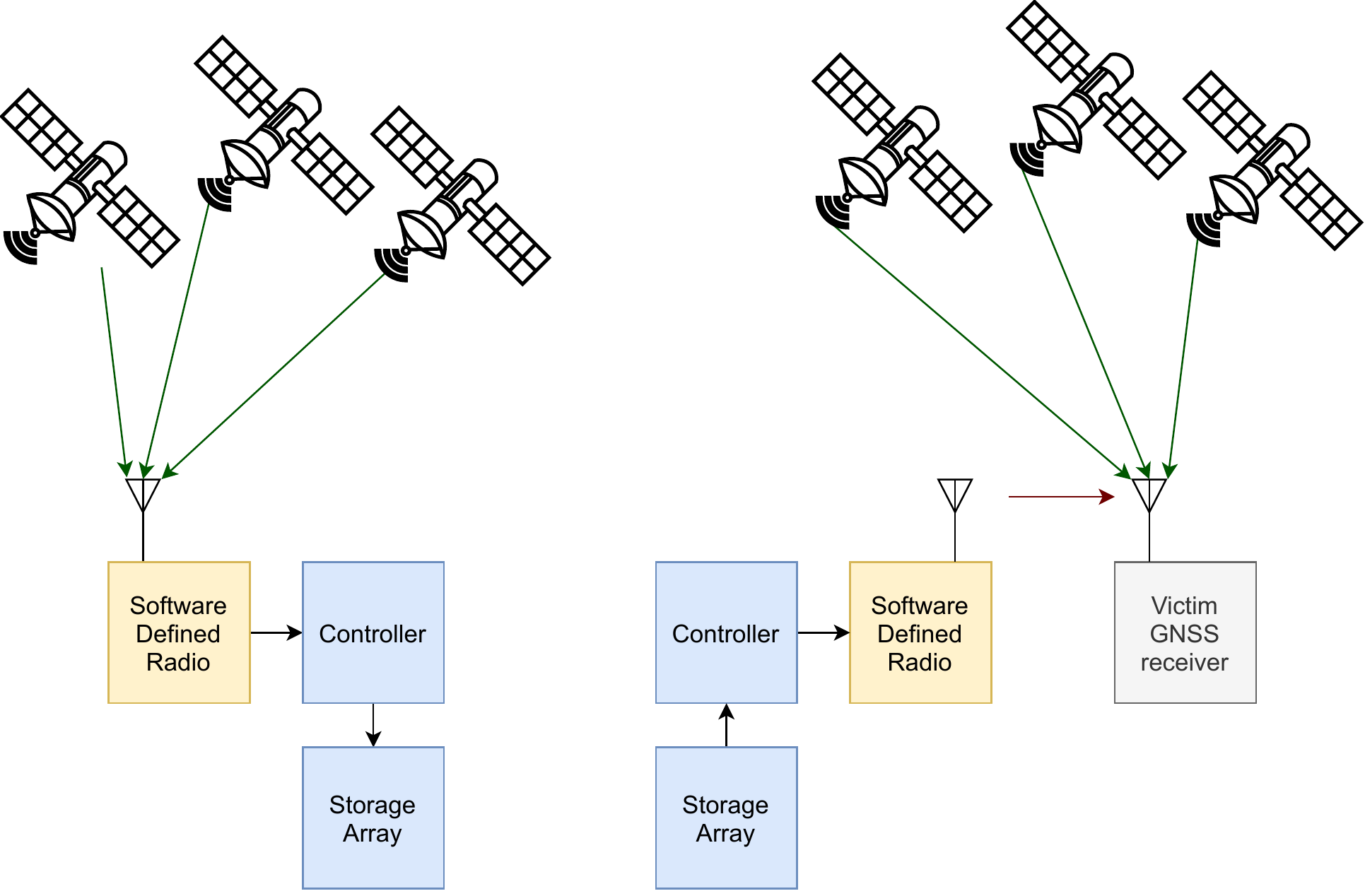}
    \caption{\rar{}: the attacker replays real GNSS signals at a different location or time.}
    \label{fig:record-replay-attack}
  \end{subfigure}
  \hfill%
  \begin{subfigure}[t]{0.4\textwidth}
    \includegraphics[width=\linewidth]{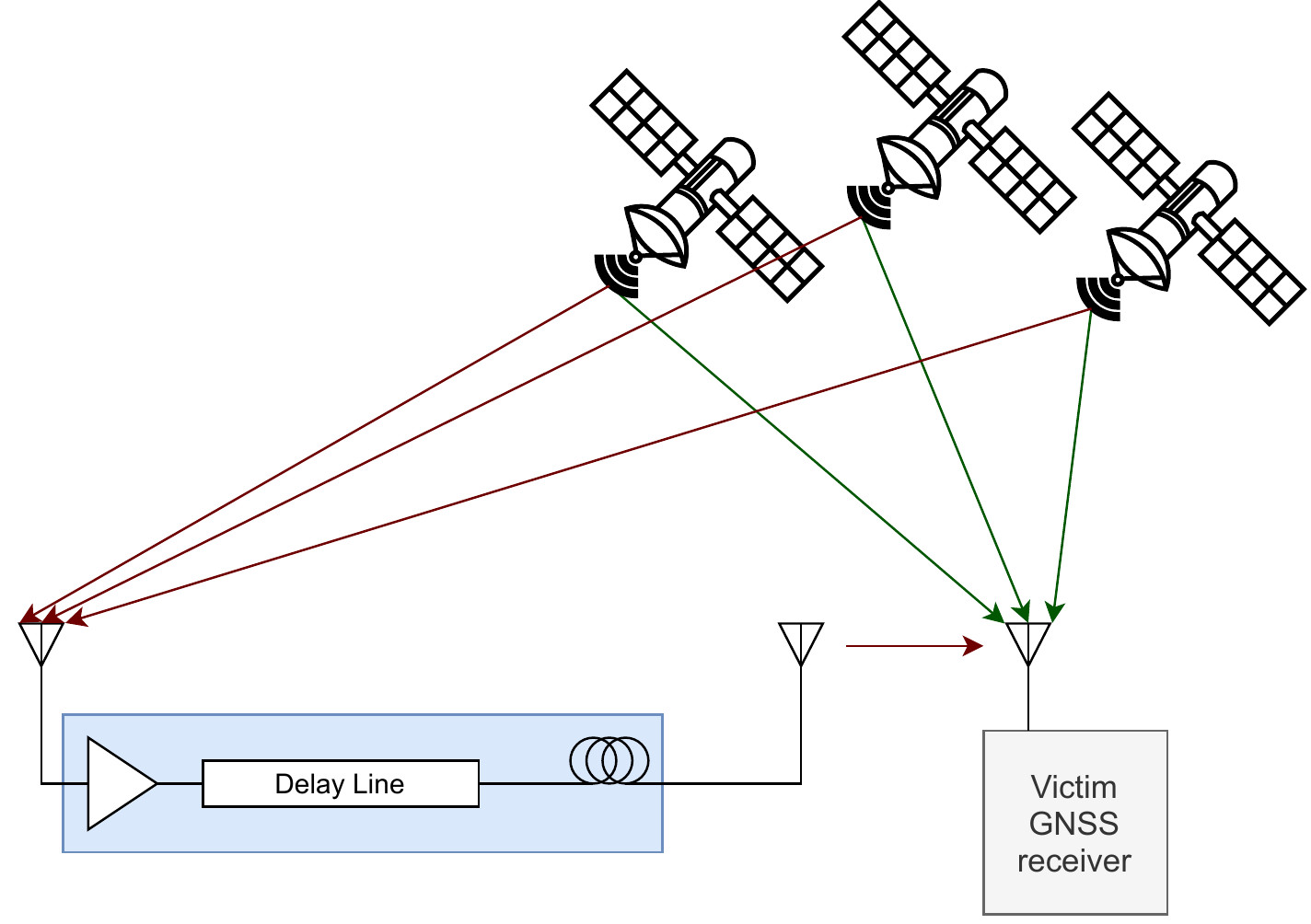}
    \caption{Meaconing: the attacker delays GNSS signals in real-time.}
    \label{fig:meaconing-attack}
  \end{subfigure}
  \caption{Different types of simplistic replay attack.}
   \label{fig:clean-static}
\end{figure*}

Existing work on replaying \gls{gnss} signals on the physical-layer mainly focuses on two attack types: \rar{}, depicted in \cref{fig:record-replay-attack}, and meaconing, i.e., delayed retransmission in real-time, depicted in \cref{fig:meaconing-attack}. 
\rar{} transceivers allow replicating realistic signal conditions in a controlled environment \cite{Ilie2011, Hickling2013a}. From an attacker's perspective, this technique can be used to replay a specific \gls{pvt} solution at a later point in time and, if desired, at a different position. 
An attacker can 'capture' a victim receiver by jamming and replaying with sufficient signal power advantage \cite{teunissenSpringerHandbookGlobal2017}. This is possible even when the victim is able to track satellites in different frequencies (e.g., simultaneous tracking of GPS L1CA and L2CA) if the attacker replays all relevant bands, by employing wide-band front-ends or coherently sampling multiple bands. The \rar{} approach is extensively demonstrated in literature \cite{Papadimitratos2008c, chenPracticingRecordandreplaySystem2013, Blum2019, brownOpenSourceSoftware2013, papadimitratosProtectionFundamentalVulnerability2008} and is also applicable in the case of authenticated signals \cite{Maier2018a, Caparra2017, Humphreys2013}. 

However, replaying systems provide only limited control on the receiver's \gls{pvt} solution. Furthermore, such systems generate up to \SI{900}{\giga \byte/\hour} for a sampling frequency of \SI{62.5}{\mega S/\second}, sampling a single channel, requiring high speed disk arrays \cite{Ilie2011}. In addition, complex digitizers and front-ends are needed to guarantee low distortion of the sampled signals. These requirements can be relaxed by trading off signal quality of the recorded signal: narrow-band recording systems distort the signal characteristics, limiting the replay of second order effects. 

Sampling frequencies as low as \SI{1}{\mega S/\s}, however, suffice for successful \rar{} attacks, when combined with brief jamming intervals \cite{Lenhart2021}. Also, 
storage requirements can be lifted if the recordings are streamed and replayed in real-time between inter-networked attacker nodes \cite{Lenhart2021}.
Such a setup can be used to replay \gls{gnss} signals over long distances, but is in practice limited by the available network bandwidth. 

For authenticated signals, an attacker has to predict the security codes if trying to match the legitimate signal at chip level in real-time. If performed successfully, this attack is very effective and more flexible than a signal level replay. This \gls{scer} attack, described in \cite{Humphreys2013}, allows an attacker to overtake a security-enhanced victim receiver, but it requires knowledge of the victim receiver's initial state. Moreover, the attack complexity is beyond the capabilities of common attackers.

\section{Adversary model}
\begin{figure*}[htb]
  \centering
  \begin{subfigure}[t]{0.42\textwidth}
    \includegraphics[width=\linewidth]{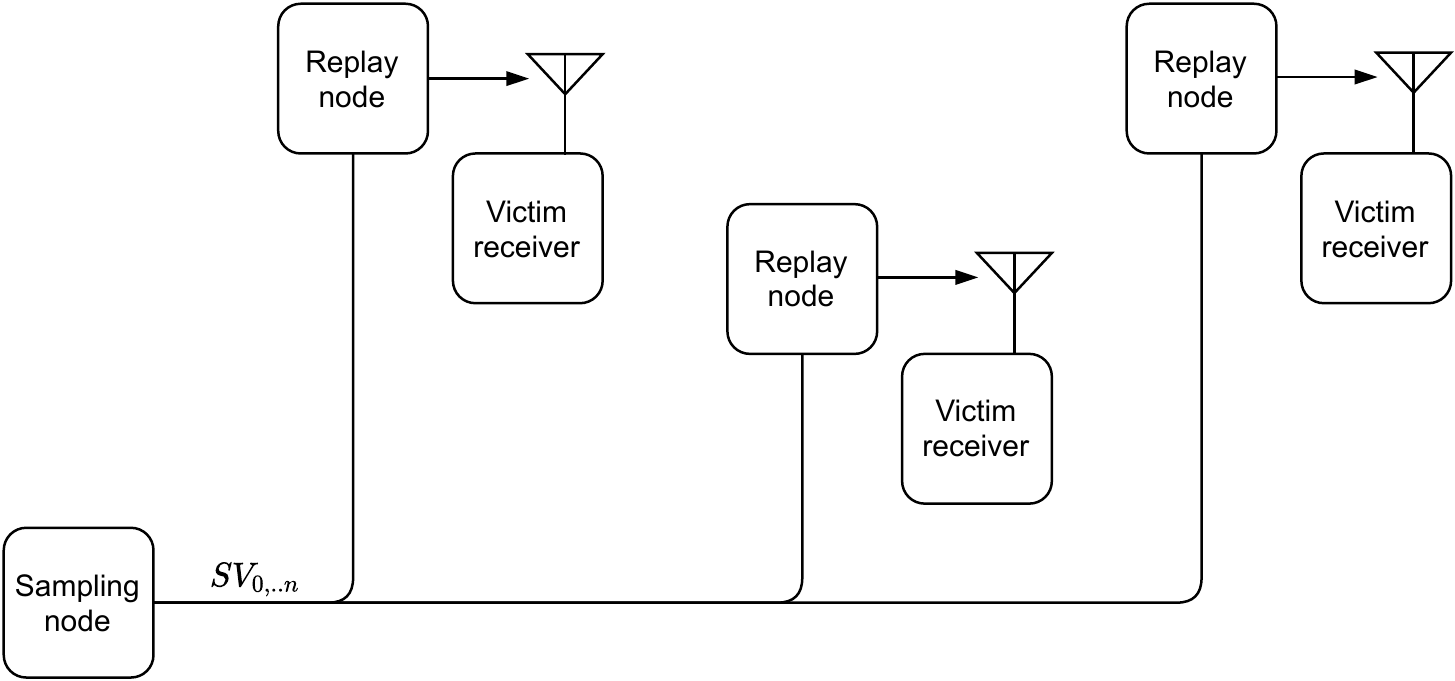}
    \caption{Distributed 1-to-many attack: each \gls{atx} node replays the received spectrum to capture a different victim receiver.}
    \label{fig:distributed-1-1-replay}
  \end{subfigure}
  \hfill%
  \begin{subfigure}[t]{0.48\textwidth}
    \includegraphics[width=\linewidth]{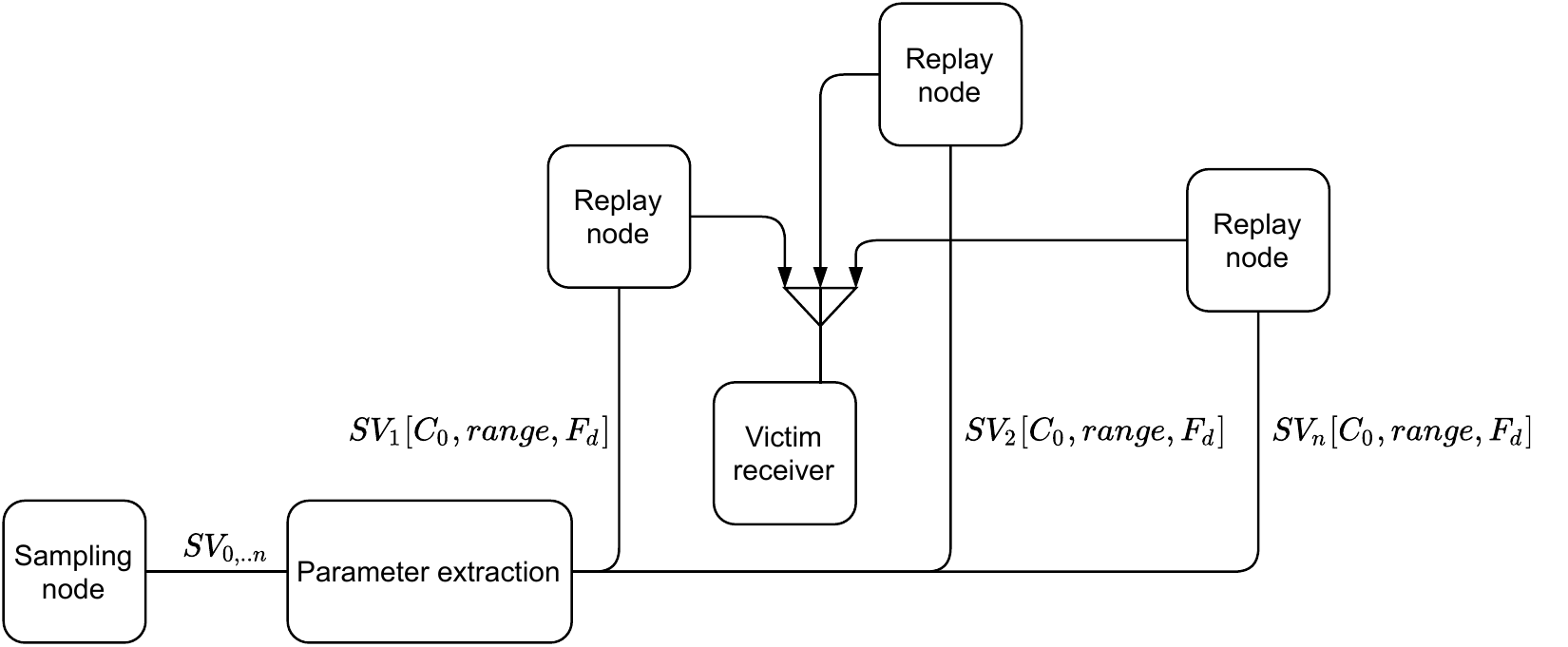}
    \caption{Distributed 1-to-many colluding attack: multiple synchronized nodes replay different subsets of the available satellites to a single victim. This attack configuration can potentially defeat \gls{aoa} spoofing detection.}
    \label{fig:distributed-many-attack}
  \end{subfigure}
  \caption{Different possible distributed replaying/relaying attack setups.}
\end{figure*}

Distributed replaying adversaries can overcome range boundaries imposed by the physical antenna cables length in classic meaconing, but they are less robust due to external factors such as network availability.
Provided that the attacker has access to network connectivity, it is possible to split the adversarial replaying device into two colluding nodes, typically operated by the same adversary. In particular, an attacker can sample \gls{gnss} signals in one position and relay these to a replaying node located in a different position, where the signals are replayed towards a targeted victim.
In order to setup such a successful real-time signal level replay, the attacker needs to deploy two or more interconnected colluding nodes with \gls{gnss} radio front-ends (\cref{fig:distributed-1-1-replay}). 
Although effective and relatively straightforward to setup, signal level replay/relay is limited by external factors (e.g., network connection quality) and it is further limited by the the amount of data exchanged between nodes. Such an attacker has to replay the entirety of the selected spectrum, without being able to filter specific satellites. Thus, it can potentially be defeated by antenna-based countermeasures that determine the \gls{aoa} of received satellite signals. 
Deploying multiple \gls{atx} nodes does not help the signal level replayer, as all satellites would suspiciously appear to originate from multiple sources, beyond expected multipath effects.

To overcome this limitation, we enhance the basic meaconing attacker. First, the \gls{arx} node processes the available \gls{gnss} signals in real-time to obtain navigational information, authentication bits, signal properties, and satellite messages.  
To significantly reduce network utilization, only these relevant parameters regarding the \gls{gnss} signals are relayed to the \gls{atx} nodes. 
At the \gls{atx} node, \gls{gnss} signals based on relayed signal- and message-properties are re-generated.
This approach adds latency compared to replaying on signal level, but it enables a much broader set of attacks. Both approaches work on \gls{gnss} authenticated signals, but not on fully encrypted signals, as demodulation and parameter extraction is not possible without access to the secret spreading codes.

When operating on message level, an attacker can further operate in a more synchronized and distributed configuration, as depicted in \cref{fig:distributed-many-attack}.
By demodulating individual satellite signals, she is able to distribute a chosen subset of satellite signals to one or multiple \gls{atx} nodes in the proximity of the victim receiver. 
The victim receiver(s) obtain(s) the superposition of the re-generated attacker signals, together with the weaker, legitimate \gls{gnss} signals. Such a distributed attacker can reconfigure the satellite assignment and distribution at run-time, thus compensating network congestion, node failure (or capture) or other attack degrading factors. This orchestrated attack approach potentially increases attack effectiveness against \gls{aoa} detection.

Due to the achieved compression by operating at message level, requirements on the connectivity between the \gls{arx} and \gls{atx} nodes can be relaxed. In fact, assuming that each node can access cellular infrastructure, this attack is suited for adversarial nodes connected to mobile cellular networks. The distributed adversary is capable of positioning the \gls{arx} node close to the victim receiver. 
By using highly directional antennas and proper shielding, the attacker can avoid self spoofing effects. This allows the attacker to use legitimate \gls{gnss} signals for time-synchronization of the colluding nodes.
After capturing the victim receiver by jamming and overpowering, the attacker has full control on the victim's \gls{pvt} solution, either by introducing selective delays to satellite streams, and/or by initiating movement with the adversarial \gls{arx} node. In a more advanced setting, the attacker can position its \gls{arx} node close to the true victim position initially. This avoids sharp discontinuities in the victim \gls{pvt} solution. 

In the scope of replay attacks, we distinguish receivers operating in cold start mode (i.e. not locked to legitimate satellites) and those which already obtained a \gls{pvt} fix. 
After a cold start, the receiver performs a search of satellites in view, called acquisition. %
During acquisition, a receiver is relatively easy to spoof, as it will lock onto the strongest satellite signals, which, due to proximity advantage, are those of the attacker. In contrast, a receiver already locked to legitimate signals keeps tracking these, unless the attacker forces a loss of lock by jamming, before initiating the spoofing attack.
As victim receiver we consider an advanced general purpose \gls{cots} multi-\gls{gnss} receiver that features anti-jamming and anti-spoofing capabilities. Additionally, we consider both static and mobile spoofing scenarios in different environments (i.e. open sky and urban setting). 
\label{sec:attacker-model}

\section{Implementation}
\label{sec:implementation}
To demonstrate the effectiveness of the advanced distributed relay/replay attacker, we develop a mobile test bench with \gls{arx} and \gls{atx} nodes built from \gls{cots} hardware. A BladeRF 2.0 \gls{sdr} is used as radio front-end to receive and transmit \gls{gnss} signals. Each node is connected to and powered by a laptop computer used to process the \gls{gnss} information. A reference \gls{gnss} receiver is used to record the true trajectory of the attacker \gls{arx} node, as well as to provide a precise clock discipline.

\begin{figure*}[htb]
  \centering
  \begin{subfigure}[t]{0.43\textwidth}
    \includegraphics[width=\linewidth]{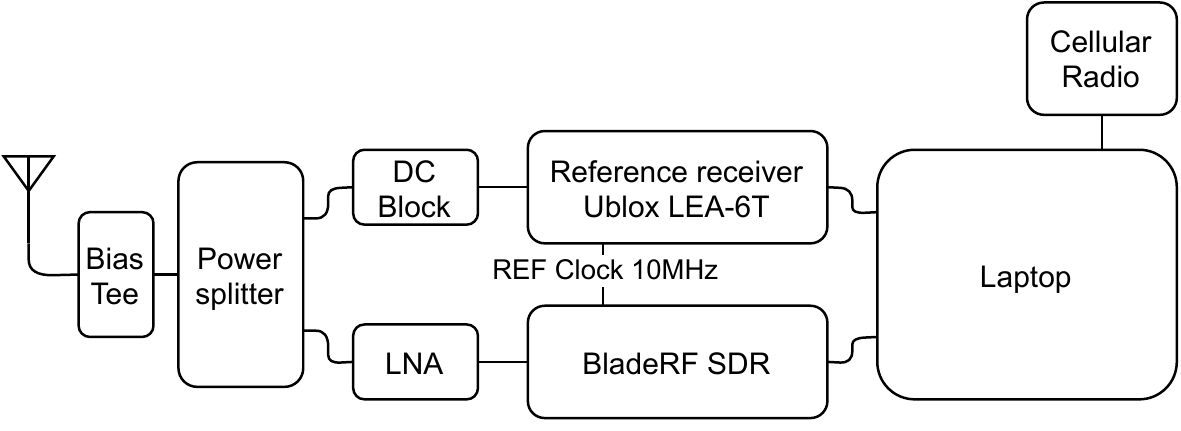}
    \caption{\gls{arx} node.}
    \label{fig:sampling-node}
  \end{subfigure}
  \hfill%
  \begin{subfigure}[t]{0.48\textwidth}
    \includegraphics[width=\linewidth]{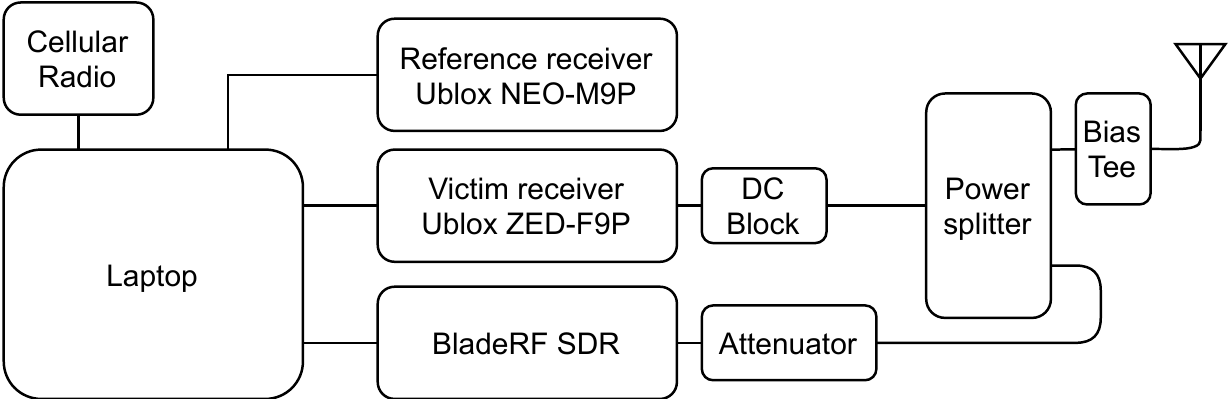}
    \caption{\gls{atx} node.}
    \label{fig:replay-node}
  \end{subfigure}
  \caption{Distributed colluding adversary block diagram.}
   \label{fig:block-diagram-devices}
\end{figure*}

\begin{figure*}[htb]
  \centering
  \begin{subfigure}[b]{0.45\textwidth}
    \includegraphics[width=\linewidth]{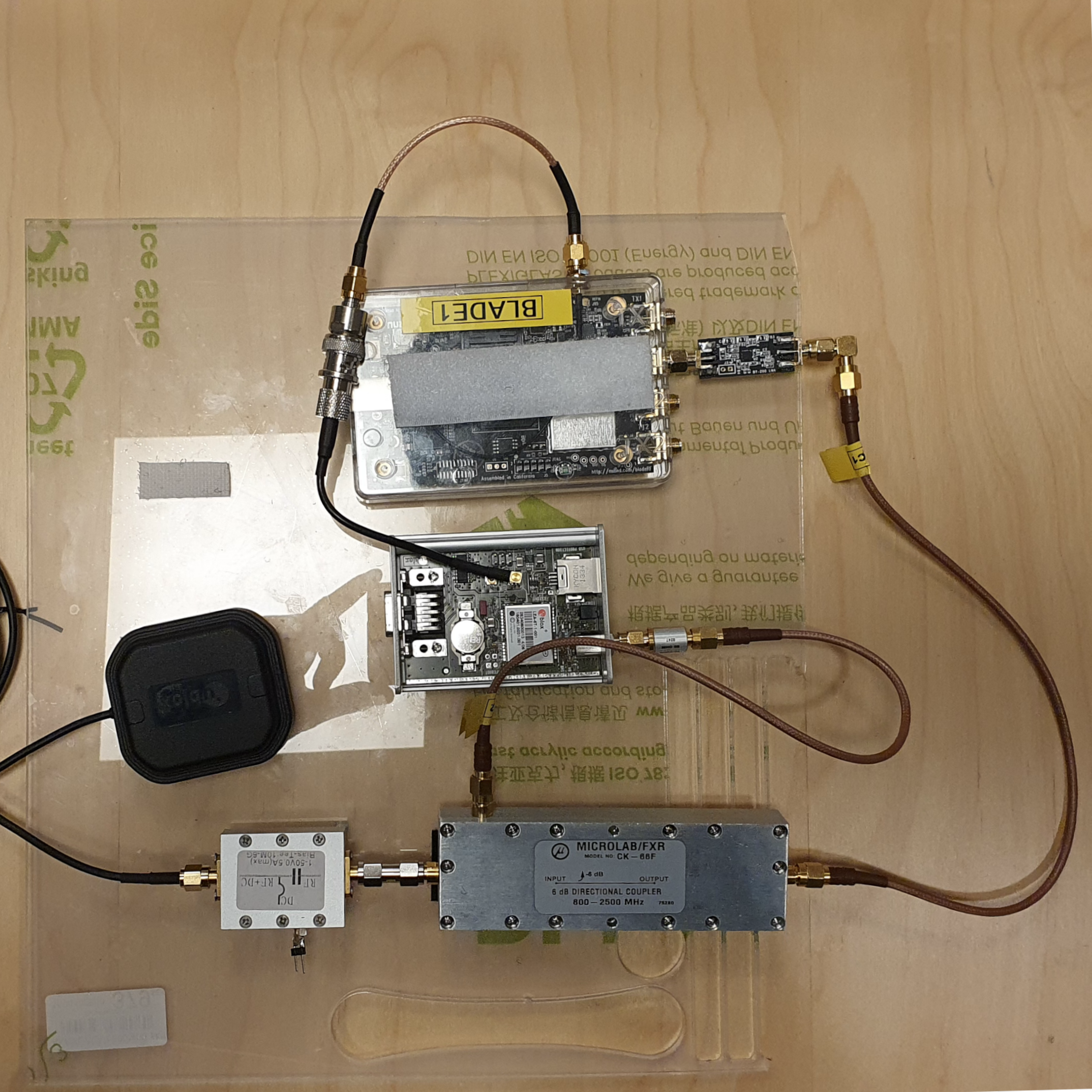}
    \caption{\gls{arx} node device. The control laptop and communication radio are not shown.}
    \label{fig:sampling-device}
  \end{subfigure}
  \hfill%
  \begin{subfigure}[b]{0.45\textwidth}
    \includegraphics[width=\linewidth]{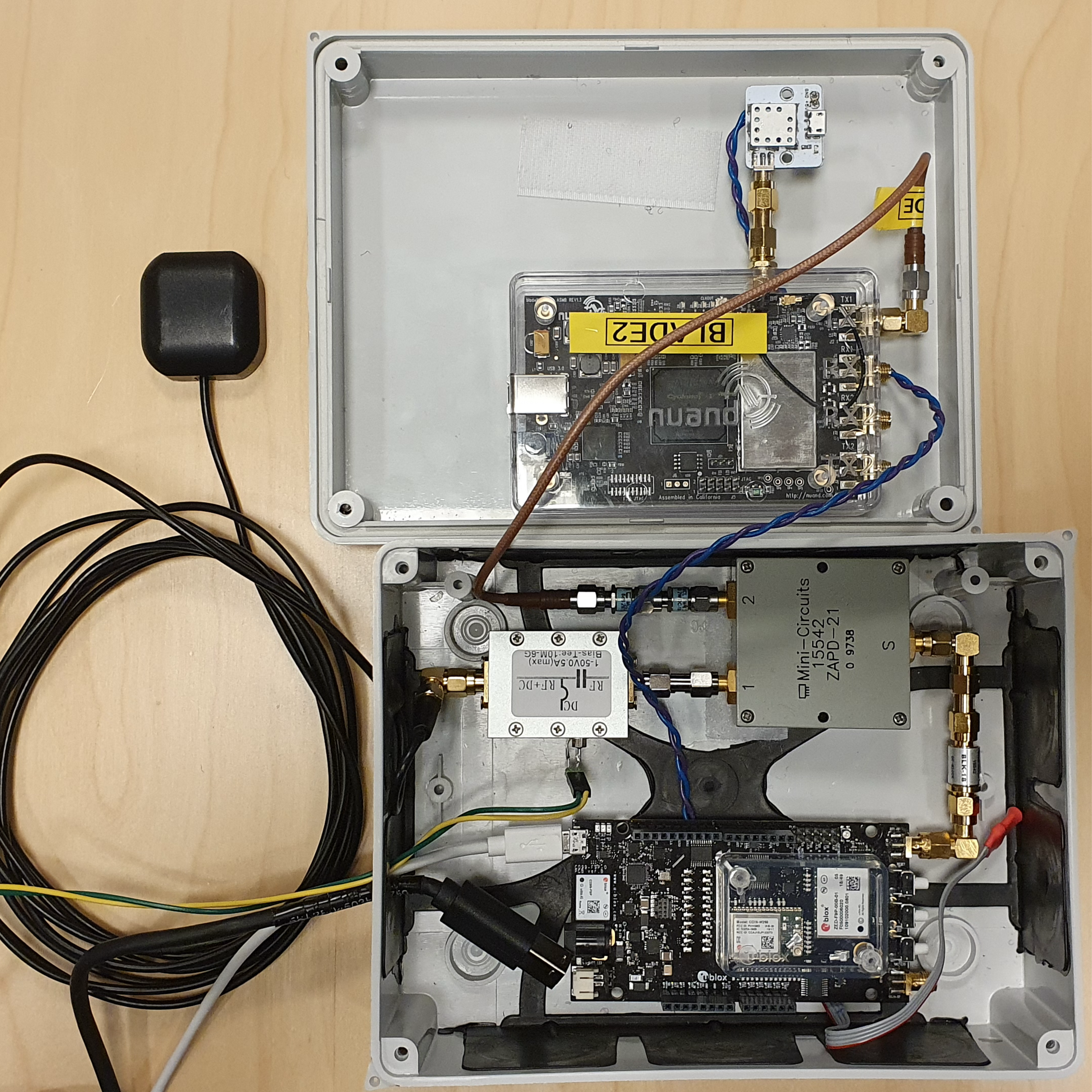}
    \caption{\gls{atx} node and victim \gls{gnss} receiver. The reference receiver, control laptop and communication radio are not shown.}    
    \label{fig:replay-device}
  \end{subfigure}
  \caption{Prototypes used for experimental evaluation of the proposed relaying/replaying schemes.}
   \label{fig:physical-devices}
 \end{figure*}

For the \gls{arx} node (\cref{fig:sampling-node,fig:sampling-device}), the signal from an active \gls{gnss} antenna is split between the reference receiver and the \gls{arx} node's \gls{sdr}. A low noise amplifier is connected in series to the \gls{sdr} front-end, to increase the sensitivity of the sampling receiver and to compensate for the signal power reduction caused by the splitter. The reference receiver (u-blox LEA-6T) is configured to output navigation and time information. A consumer grade laptop (Dell XPS 15) samples the \gls{gnss} signal in real-time, and depending on the replay mode, it extracts signal parameters. These parameters, or the raw signal samples, are relayed over an LTE connection (LTE Cat 12 with a maximum theoretical transfer speed of \SI{600}{\mega\bit/\second}), provided by the attached cellular module. Due to limitations imposed by the cellular carrier, mobile phones are not provided with a public IP, required for routing data to the \gls{atx} node. We therefore route all network traffic through a \gls{vpn} hosted at the KTH Royal Institute of Technology.

Similarly, the \gls{atx} node is connected to the same type of \gls{sdr}, cellular module and processing laptop. The signal received from the \gls{arx} node is re-transmitted via the \gls{atx} \gls{sdr}, combined with the legitimate \gls{gnss} signals and received at the victim receiver (u-blox Zed-F9P). A calibrated \gls{tcxo} is used at the \gls{atx} node's radio for precise clock discipline. Experimental observations show that this is required in cold climates for precise RF tuning. The block diagrams and utilized hardware for the \gls{atx}  are shown in \cref{fig:replay-node,fig:replay-device}.

\begin{figure}[htb]
    \includegraphics[width=\linewidth]{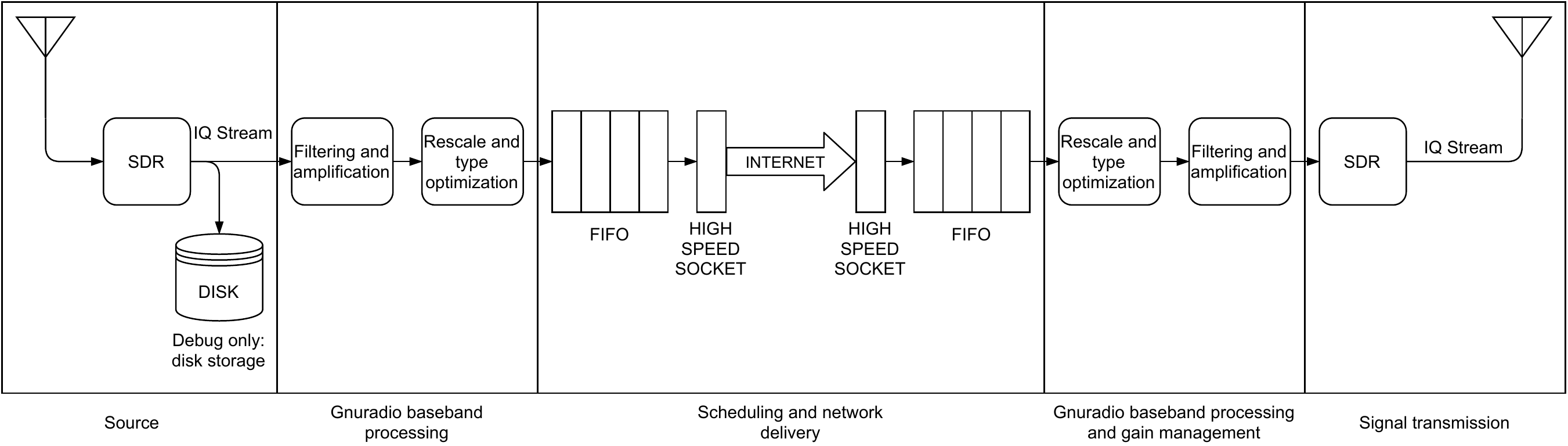}
    \caption{Signal level relaying/replaying data flow.}
    \label{fig:signal-meaconer}
\end{figure}

\textbf{Signal level replay} is performed by sampling the \gls{gnss} spectrum (in our case, we limit the replay to the \gls{gps} L1 band, due to available hardware constraints) at \SI{1}{\mega S/\second} sample rate and \SI{1}{\mega\hertz} bandwidth. \gnuradio{} \cite{gnu-radio} flows are used to handle the \glspl{sdr}, signal sampling and I/Q correction. The sampled data is transferred to the \gls{atx} node via a TCP socket. 
\cref{fig:signal-meaconer} outlines the structure of the signal level replayer. The baseband signals from the \gls{sdr} are transmitted with \SI{16}{\bit} quantization and converted back at the \gls{atx} node. A programmable gain stage provides control of the amplification level. Raw I\/Q samples obtained from the \gls{atx} node are stored to disk for analysis and validation purposes.

\begin{figure}[htb]
    \includegraphics[width=\linewidth]{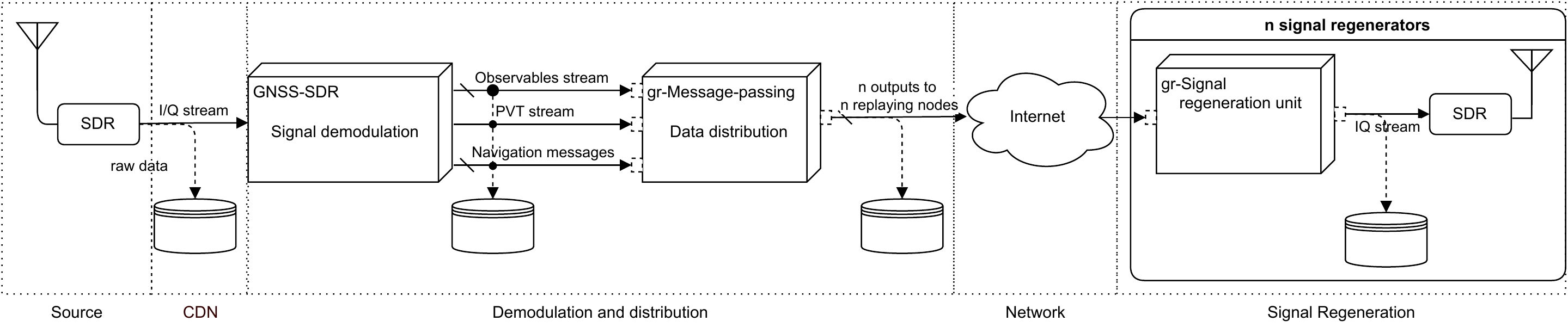}
    \caption{Message level relaying/replaying signal and message flow.}
    \label{fig:message-meaconer-flow}
  \end{figure}

\textbf{Message level replay} is performed by transmitting only relevant signal parameters, required to re-create the baseband signal at the \gls{atx} node. 
A schematic overview is depicted in \cref{fig:message-meaconer-flow}. It is built upon three existing open-source projects: \gnsssdr{} \cite{gnsssdrProc}, \gnuradio{} and \gpssdrsim{} \cite{ebinumaOsqzssGpssdrsim2020}.

The \gls{arx} node uses the open-source software \gls{gnss} receiver \gnsssdr{}. Recently added stream capabilities are used to extract useful parameters.
The transmitted signal parameters include the \gls{tow} (to synchronize multiple message streams), the \gls{pvt} position and time, send out at specific time intervals. Other parameters, such as estimates of pseudorange, Doppler frequency shift, code-phase and Carrier-to-Noise parameters for each satellite, are not yet used in this implementation phase, to not add additional measurement uncertainty. Instead, these parameters are generated by simulation. Furthermore the navigation message bits are forwarded from \gls{arx} to \gls{atx}, so that the authentication bits can be inserted into the generated navigation messages, as soon as they become perceivable for \gls{gnss} receivers.
The 'intermediate receiver spoofer' described in \cite{humphreys2008assessing} uses a similar set of signal parameters (estimates of code-phase, Doppler frequency shift and carrier-phases) and \gls{pvt} solution, to align the spoofed signals to the legitimate ones, which in turn enables advanced \gls{tsa} attacks. 
Our relaying/replaying implementation additionally forwards observed unpredictable authentication bits, which by design introduces a delay into the system and prevents synchronization to the legitimate signals to a degree necessary for \gls{tsa}. The difference in evaluated attack target, i.e., authenticated signals, determine which parameters are necessary and which attack features are achievable.

\begin{figure}[htb]
    \includegraphics[width=\linewidth]{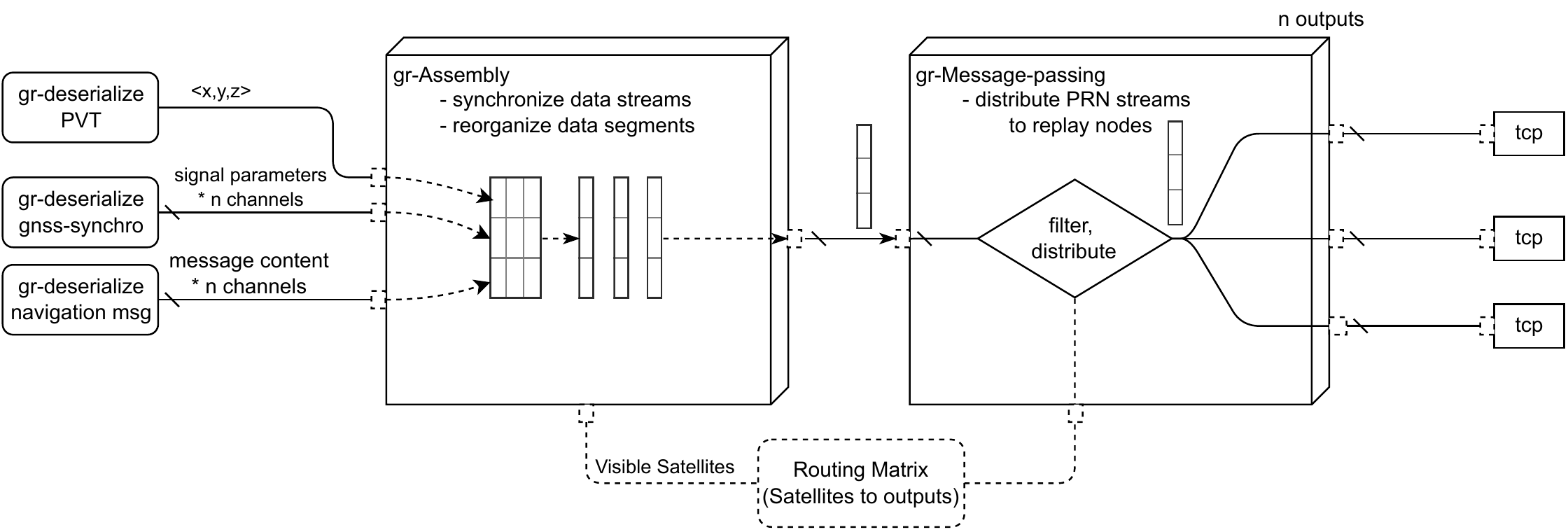}
    \caption{Message passing in \gnuradio{}: de-serialization, assembly and distribution.}
    \label{fig:message-meaconer-routing}
  \end{figure}

The \gls{arx} node flow is depicted in \cref{fig:message-meaconer-routing}. Each \gnsssdr{} output stream is received, de-serialized in its own dedicated block, from which the the contained messages are forwarded to an assembly block, synchronizing and combining the different message streams. Custom \gnuradio{} out-of-tree blocks are used for message passing, communication, as well as signal generation and streaming from and to the \glspl{sdr}. This approach helps us to the software modular, and to integrate the different replaying components easily.
A separate distribution block with knowledge of the number of \gls{atx} nodes and the respectively assigned satellites, filters the message objects, so that each node only receives the intended satellite parameters. The alternative design, in which each \gls{atx} node decides which satellites to replay, was discarded due to the higher configuration effort.

Finally, after transmission over the Internet, the signal is re-generated at the \gls{atx} node, by feeding the extracted signal parameters, navigation messages and authentication payload to an adapted version of the open-source \gls{gps} simulator \gpssdrsim{}.
Incorporating a signal simulator into a replay system sounds contradictory at first glance, but it reduces the development effort required to model Doppler-shift effects and to maintain code-phases between satellites. In the proposed implementation, the signal simulator is only used to recreate the physical-layer signal, while the underlying information is obtained from the \gls{arx} node.

\section{Experimental Setup} 
\begin{figure}[htb]
    \centering
    \hfill
    \begin{subfigure}[t]{0.27\textwidth}
    \centering
    \includegraphics[width=\linewidth]{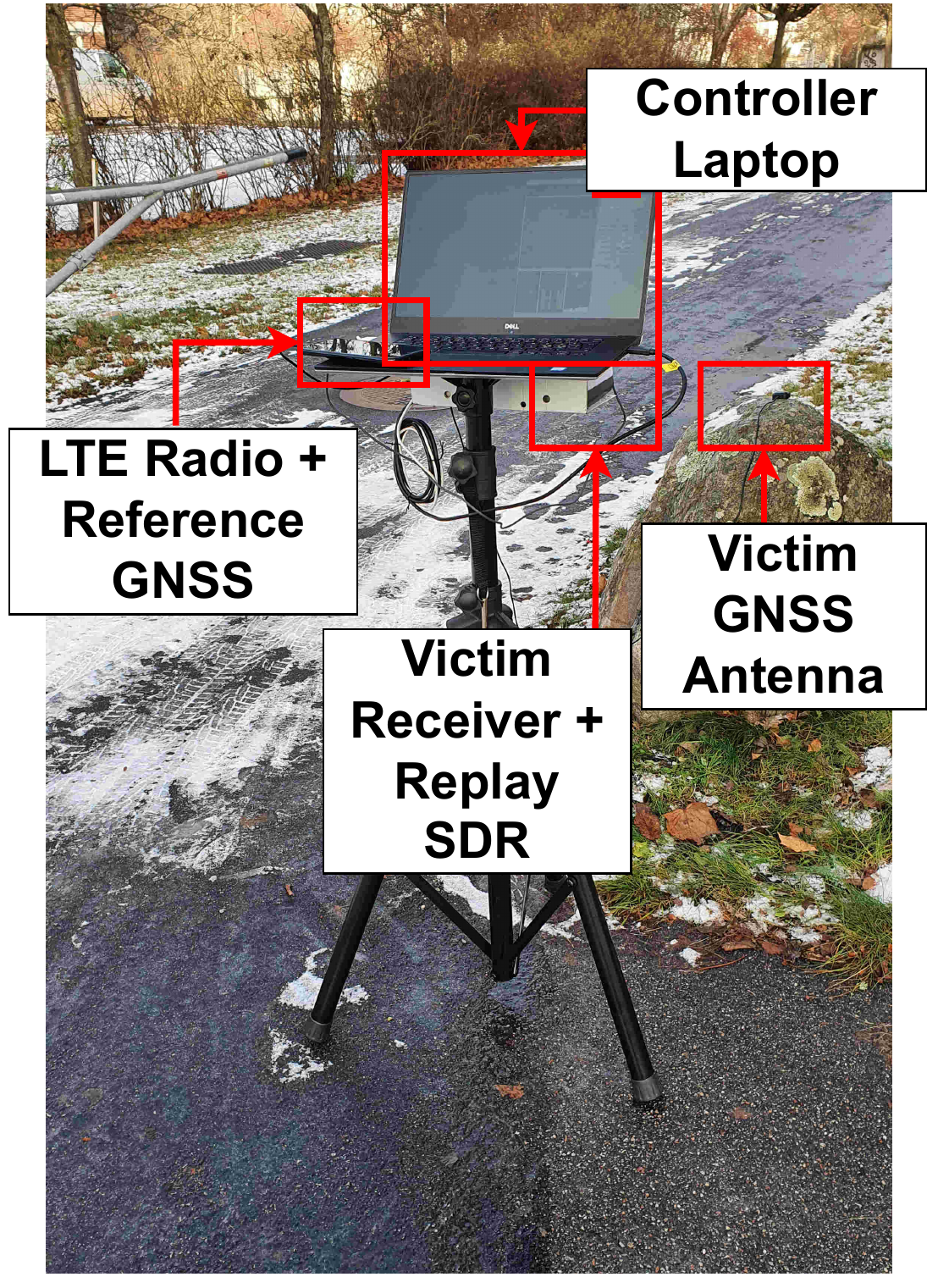}
    \caption{Victim GNSS receiver: the victim receiver is located at a fixed position and has a lock to the GNSS satellites.}
    \label{fig:victim-setup}
    \end{subfigure}
    \hfill
    \begin{subfigure}[t]{0.55\textwidth}
    \centering
    \includegraphics[width=\linewidth]{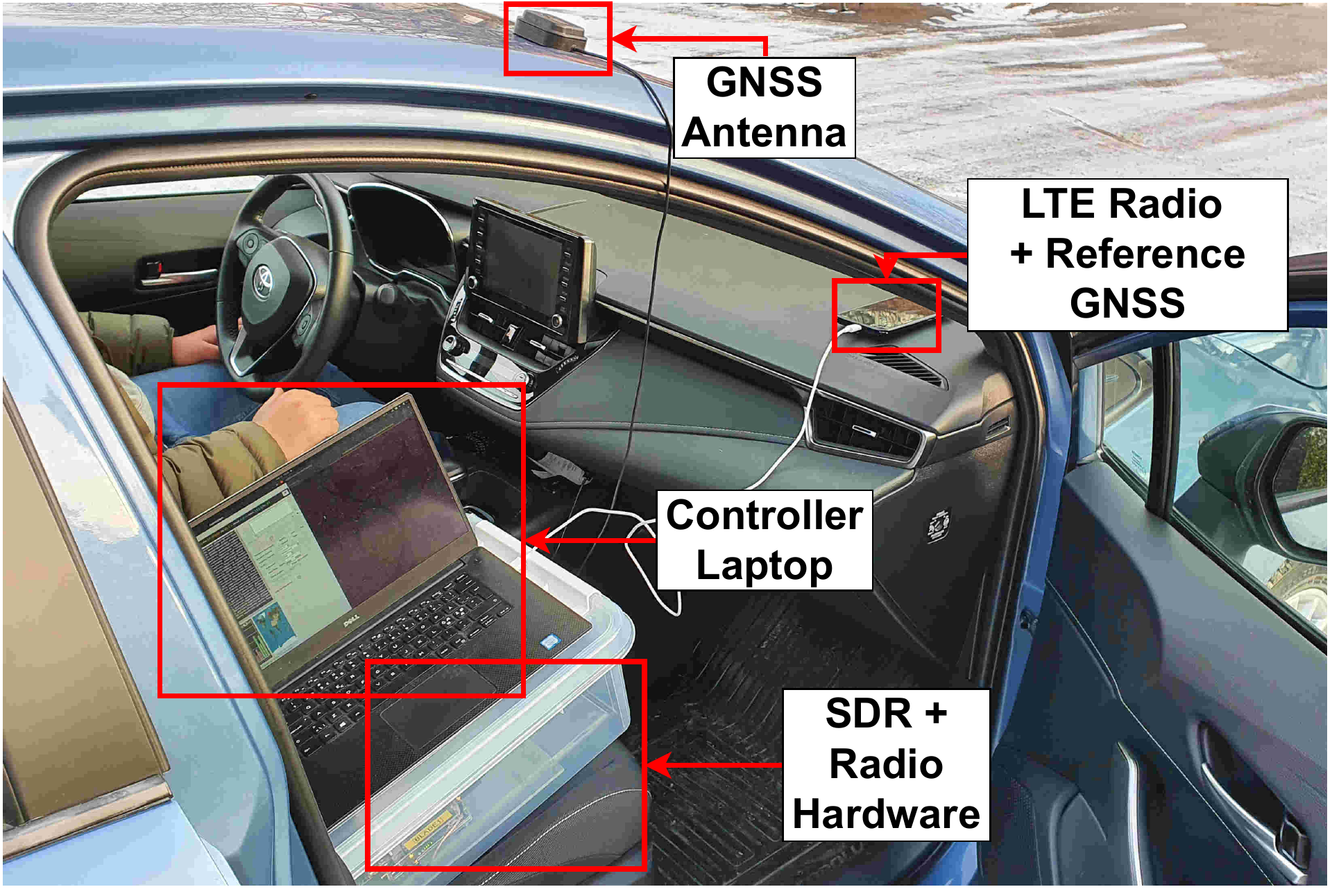}
    \caption{Attacker setup: a moving \gls{arx} node on a car collects GNSS information that is replayed at the victim.}
    \label{fig:attacker-setup}
    \end{subfigure}
    \hfill
    \caption{Experimental setup for attacking a static victim receiver using a moving \gls{arx} node.}
    \label{fig:experimental-setup}
  \end{figure}

The victim \gls{gnss} receiver is located in a fixed position $LOC_{start}$ (\cref{fig:victim-setup}). The \gls{atx} node is placed near the victim, initially passive. It is connected to the victim receiver via a power combiner and RF cables to adhere to regulatory restrictions regarding protected frequency bands.
The mobile \gls{arx} node is mounted on a vehicle (\cref{fig:attacker-setup}). Initially, it is near the victim as well, so that for all nodes and the victim receiver the ground truth location at $t=t_0$ is $LOC_{start3D}=[17.957016E;59.402846N$]. Both the \gls{arx} node and the victim are monitored using a reference \gls{gnss} receiver throughout the experiment. Similarly, the network throughput between attacker nodes is measured for the duration of the attack.

At the start of each experiment, the victim receiver and \gls{arx} node start from cold-start mode and start by acquiring and tracking satellites to derive a legitimate \gls{pvt} solution. 
After few minutes, at $t_j=t_0 + 200s$, the attacker jams the victim receiver to cause a loss of lock on the legitimate satellites. Successful jamming can be achieved in approximately \SI{15}{\second}, so that, at $t_r=t_0 + 215s$, the attacker starts replaying \gls{gnss} signals. These are directly streamed from the \gls{arx} node in the case of signal level replay, for message level replay signal parameters are extracted before streaming. 
Due to the higher \gls{snr} of the replayed signals, the victim receiver will, with high likelihood, track the adversarial signals.
We observed that at approximately $t_{adv3D} = t_0 + 300s$, the victim receiver usually derives a new \gls{pvt} fix corresponding to the attacker's signals.
Once the victim locks on the replayed signal, the \gls{arx} node starts moving physically: this, essentially, generates a spoofed \gls{pvt} trajectory, so that the victim appears to be moving alongside the \gls{arx} node.

 \begin{figure}[htb]
  \begin{subfigure}[t]{0.48\textwidth}
    \centering
    \includegraphics[height=7cm]{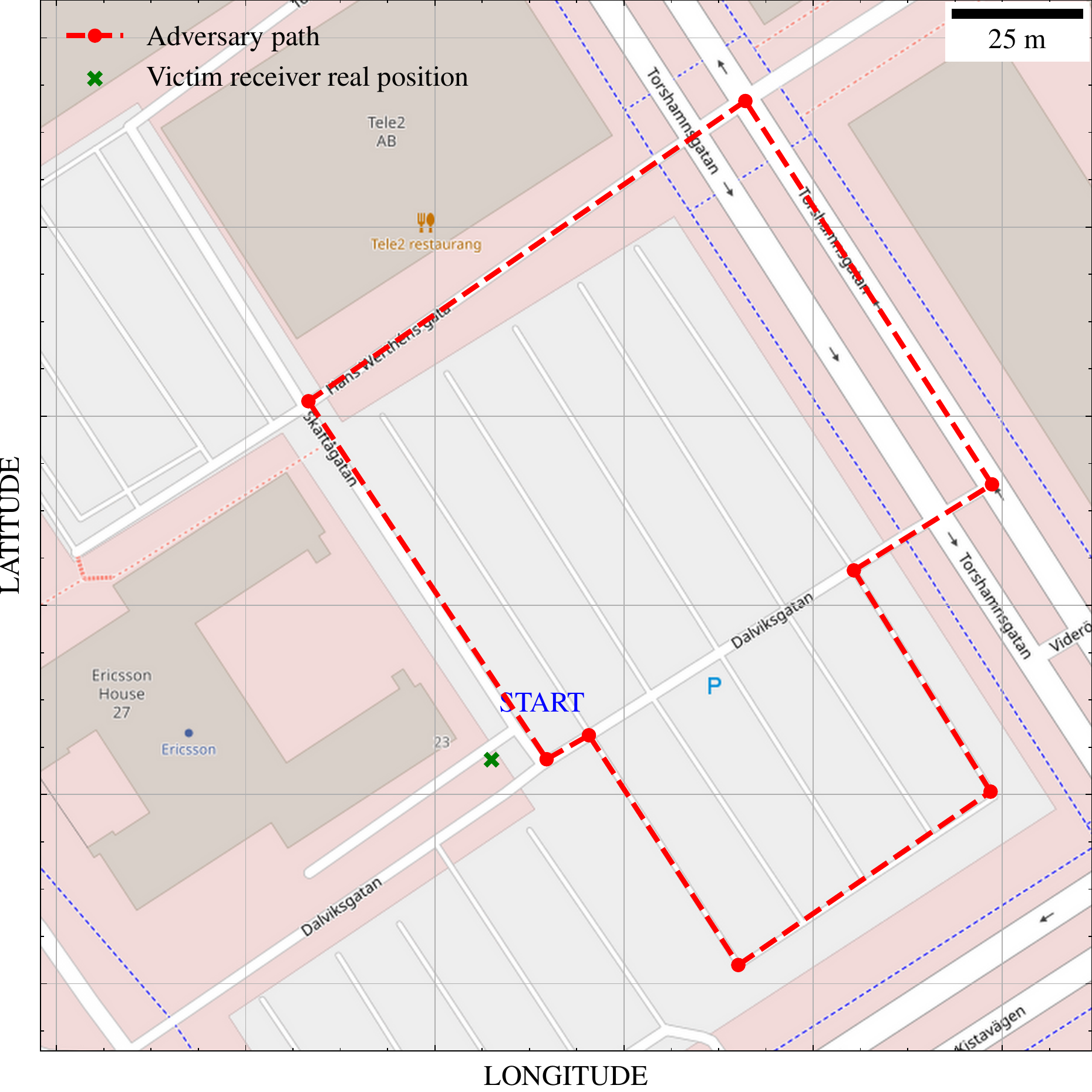}
    \caption{Ground truth for attack: the adversary moves along the path, while the victim is static at the reference position.}
    \label{fig:ground-truth-sig}
  \end{subfigure}
  \begin{subfigure}[t]{0.48\textwidth}
    \centering
    \includegraphics[height=7cm]{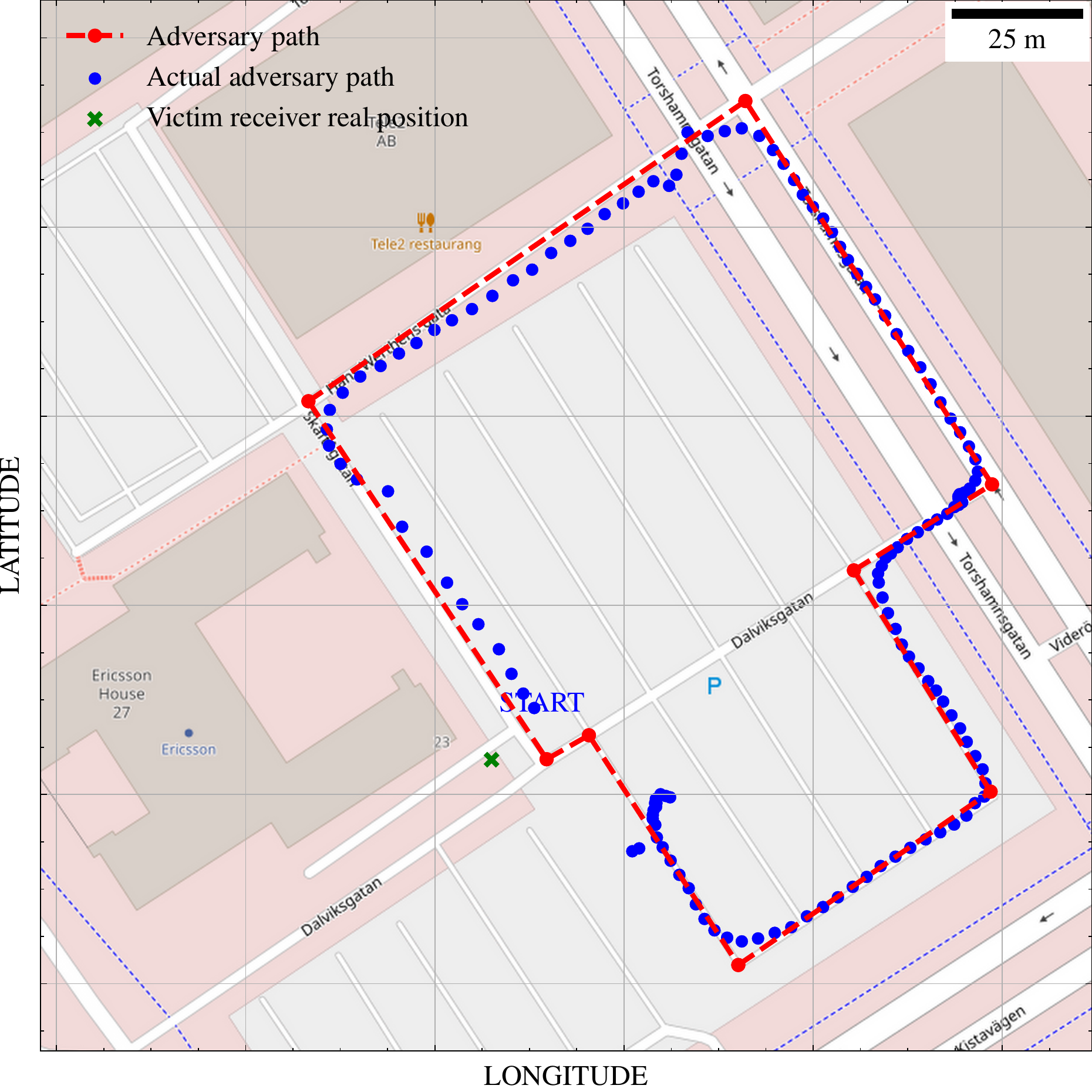}
    \caption{Sampled positions used for the message level relay/replay attack: the same ground truth path is used.}
    \label{fig:ground-truth-ml}
  \end{subfigure}
\caption{Ground truth traces for signal and message level relaying/replay.}
\label{fig:ground-truth}
\end{figure}

\cref{fig:ground-truth} shows the predefined trajectory used for the relay/replay attack in this experiment. We perform replay experiments in an urban environment and rely on cellular 4G networks to transmit the replay signal between the adversarial sampler and the transmitter. All experiments are conducted in the surrounding area of the KTH Royal Institute of Technology campus in Kista (Stockholm). The location used for testing is situated in a position that offers favorable 4G connection, with coverage offered by Tele2 Sweden network. 
\label{sec:experimental}
    
\section{Evaluation}
\label{sec:evaluation}
\begin{figure}[htb]
    \centering
    \includegraphics[width=0.6\textwidth]{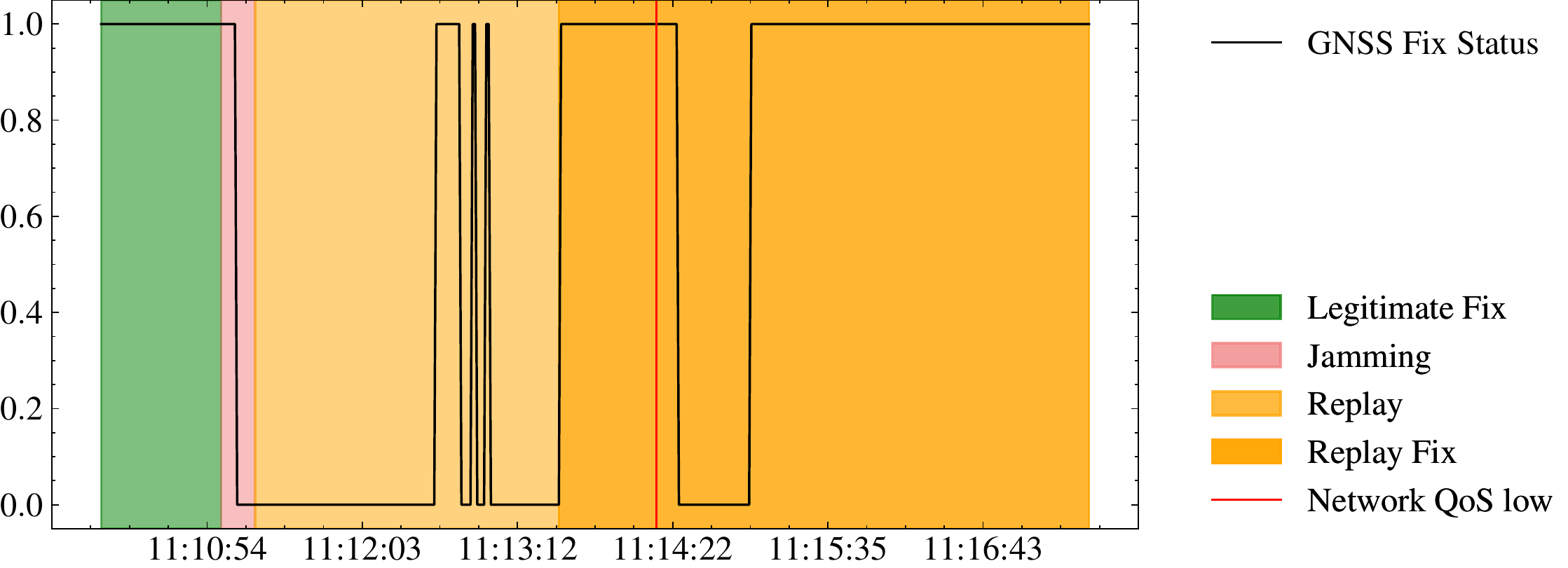}
    \caption{Phases of the attack. After the initial fix on legitimate signals: jamming, replay of attack signals, lock on replayed signals.}
    \label{fig:meacon-phases}
  \end{figure}

\begin{figure}[htb]
    \centering
    \begin{subfigure}{0.48\textwidth}
        \includegraphics[height=7cm]{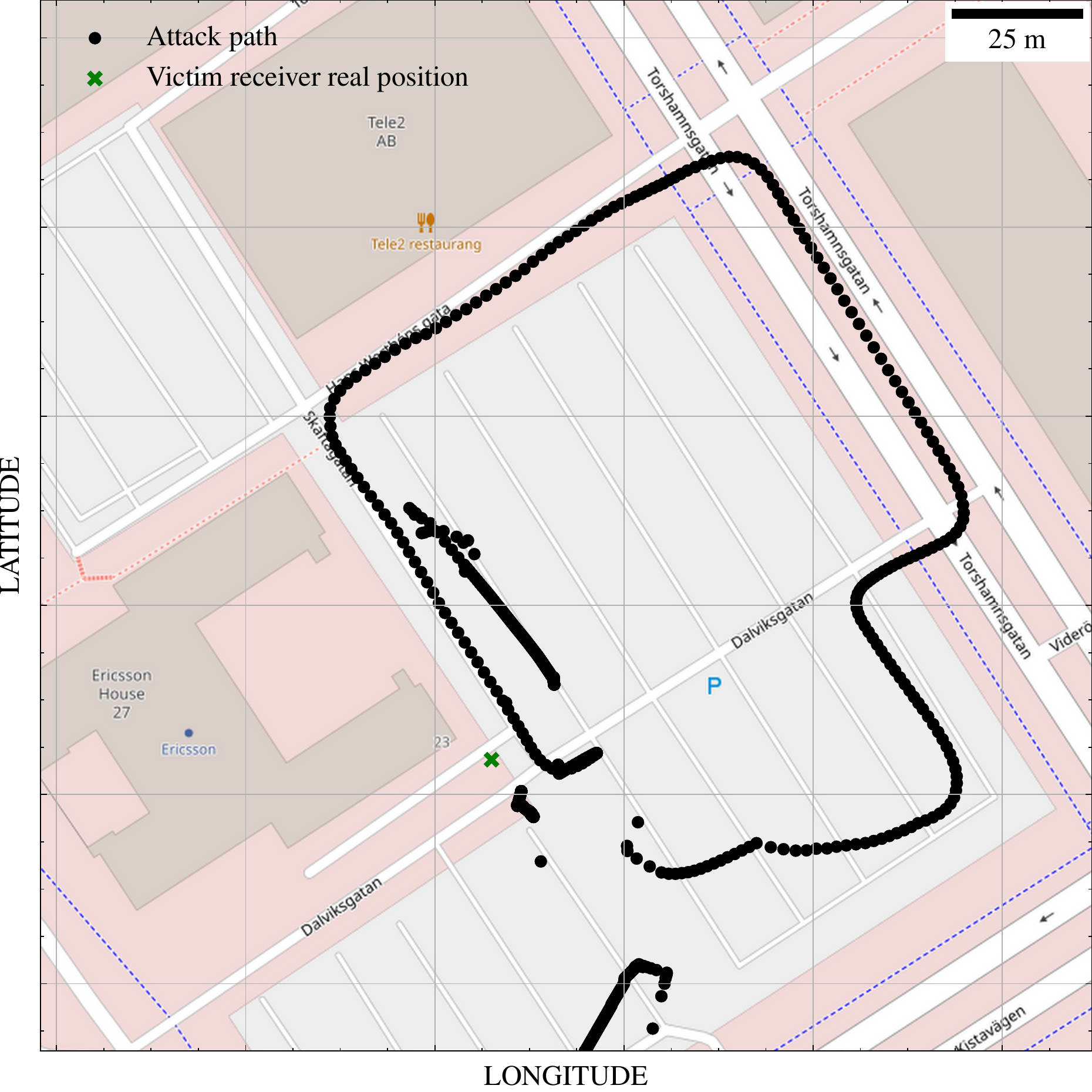}
        \caption{Attacker modified position compared to actual position of the victim.}
        \label{fig:success-ref}
    \end{subfigure}
    \begin{subfigure}{0.48\textwidth}
        \includegraphics[height=7cm]{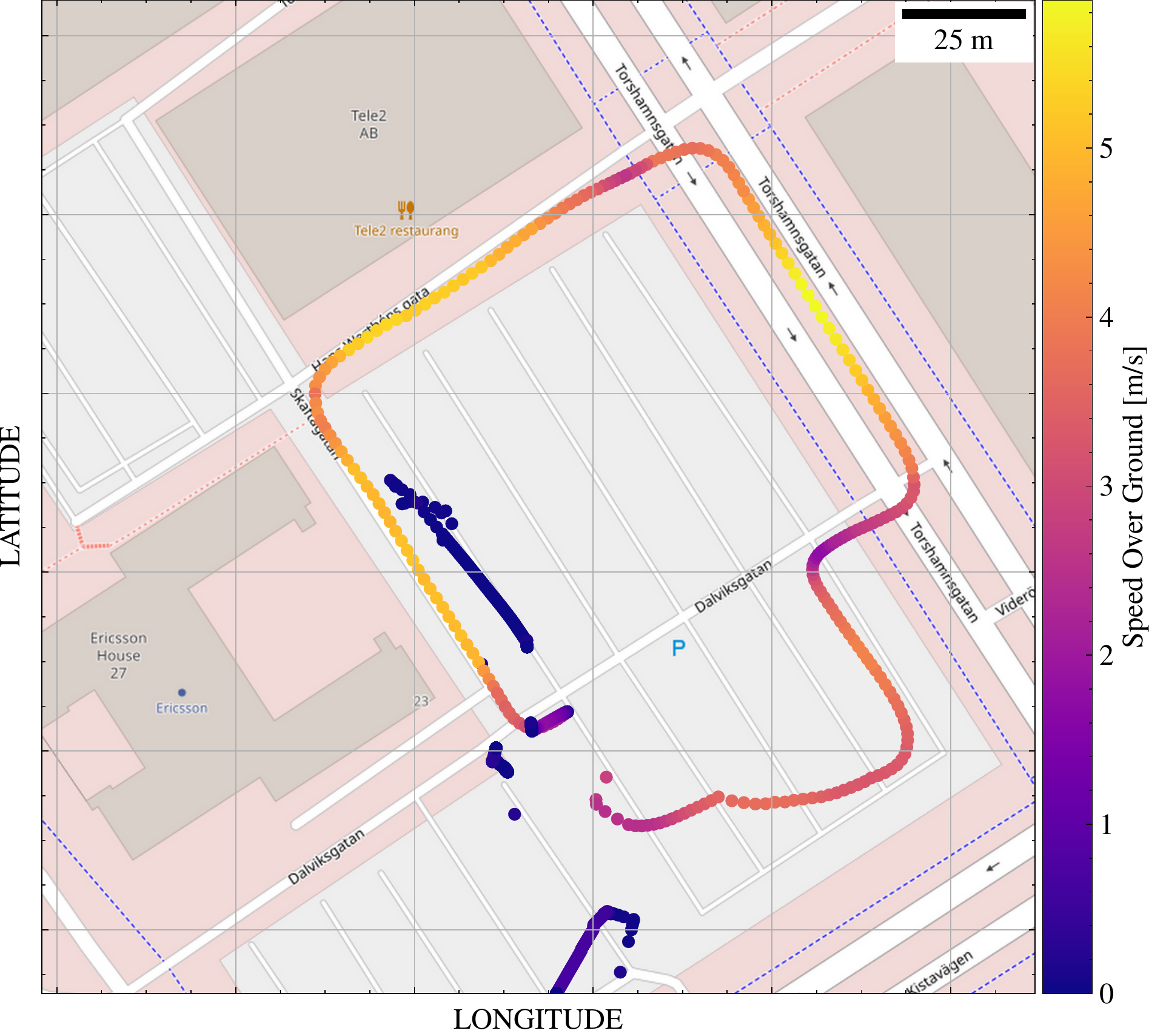}
        \caption{Attacker modified velocity.}
        \label{fig:success-velocity}
    \end{subfigure}
    \caption{Successful signal level relay/replay: the PVT solution at the victim follows the adversary movement.}
    \label{fig:success-meacon}
\end{figure}

\textbf{Signal level relaying/replaying} measurements show that the attacker can successfully 'capture' the victim receiver and impose a \gls{pvt} solution based on the relayed/replayed signal. \cref{fig:meacon-phases} shows the attack phases observed at the victim receiver, initially obtaining a fix based on legitimate signals. During the \SI{15}{\second} jamming period it looses lock. This is visible approximately \SI{200}{\second} into the recording. 
After the jamming phase, the replay signals overpower the legitimate signals in view. The receiver adapts to the new signals and eventually acquires a new fix, based on the initial position of the \gls{arx} node. The stability of the new fix is dependent on the ability of the attacker to consistently replay \gls{gnss} signals. Once the \gls{arx} node starts moving on the predefined path, the victim receiver seemingly follows this change in position and velocity as shown in \cref{fig:success-ref} and \cref{fig:success-velocity} respectively.

  net

\begin{figure}[htb]
  \centering
  \begin{subfigure}[t]{0.48\textwidth}
    \includegraphics[width=\linewidth]{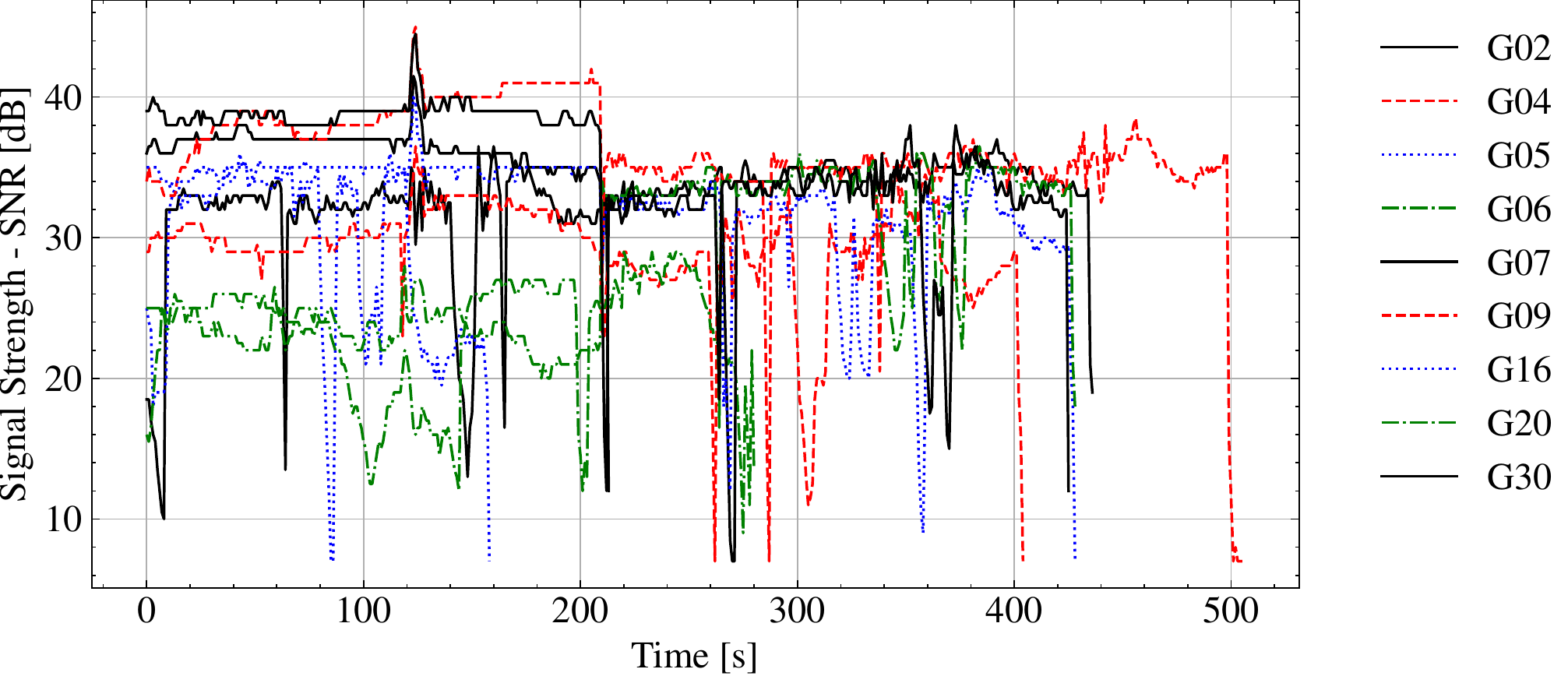}
    \caption{Victim receiver SNR.}
    \label{fig:snr-victim}
  \end{subfigure}
  \hfill%
  \begin{subfigure}[t]{0.48\textwidth}
    \includegraphics[width=\linewidth]{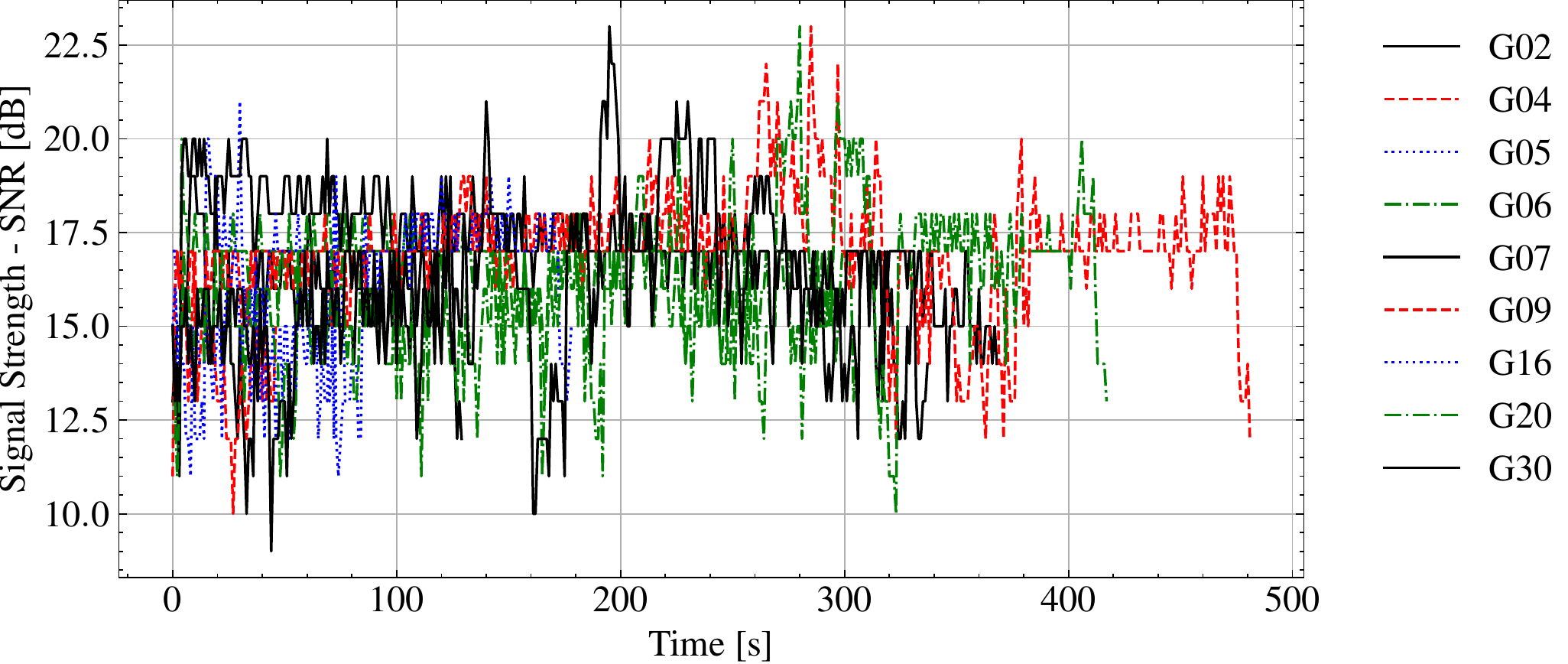}
    \caption{Reference receiver SNR at adversarial \gls{arx} node.}
    \label{fig:snr-ref-sampler}
  \end{subfigure}
  \caption{SNR raw measurements at the victim and \gls{arx} node.}
   \label{fig:snr-attack-vs-ref}
\end{figure}

Additional confirmation of the successfully attack can be obtained by observing raw signal characteristics. In particular, \gls{snr} changes correspond to the different attack phases: significant drops in signal quality during jamming and initial replay signal injection phases, as shown in \cref{fig:snr-victim}. 
The SNR bias between reference receivers at the adversarial sampler (\cref{fig:snr-ref-sampler}) and the victim (\cref{fig:snr-victim}) is due to a mismatch between the receiver's internal gain. Due to limitations of the available hardware, \gls{snr} measurements were not acquired directly at the adversarial sampler, but in a separate receiver. This receiver used a mobile phone embedded antenna located on the car's dashboard, in contrast to the sampler's high performance, high gain antenna placed on the rooftop, accounting for the \gls{snr} differences.

\begin{figure}[htb]
    \centering
    \begin{subfigure}{0.45\textwidth}
        \centering
        \includegraphics[width=0.9\linewidth]{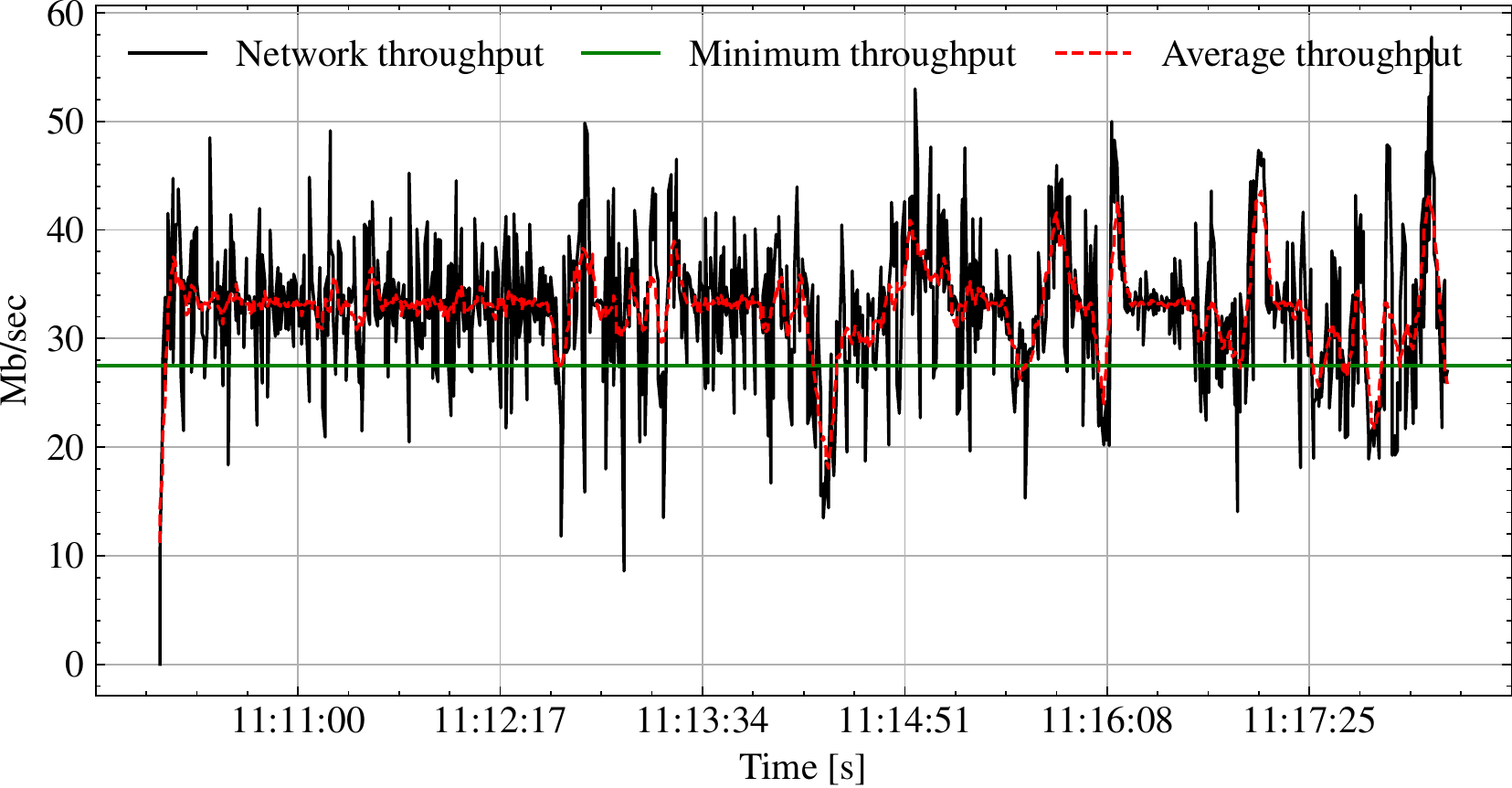}
        \caption{Signal-level network throughput measured at the \gls{arx} node.}
        \label{fig:network-log}
    \end{subfigure}
    \begin{subfigure}{0.45\textwidth}
        \centering
        \includegraphics[width=0.9\linewidth]{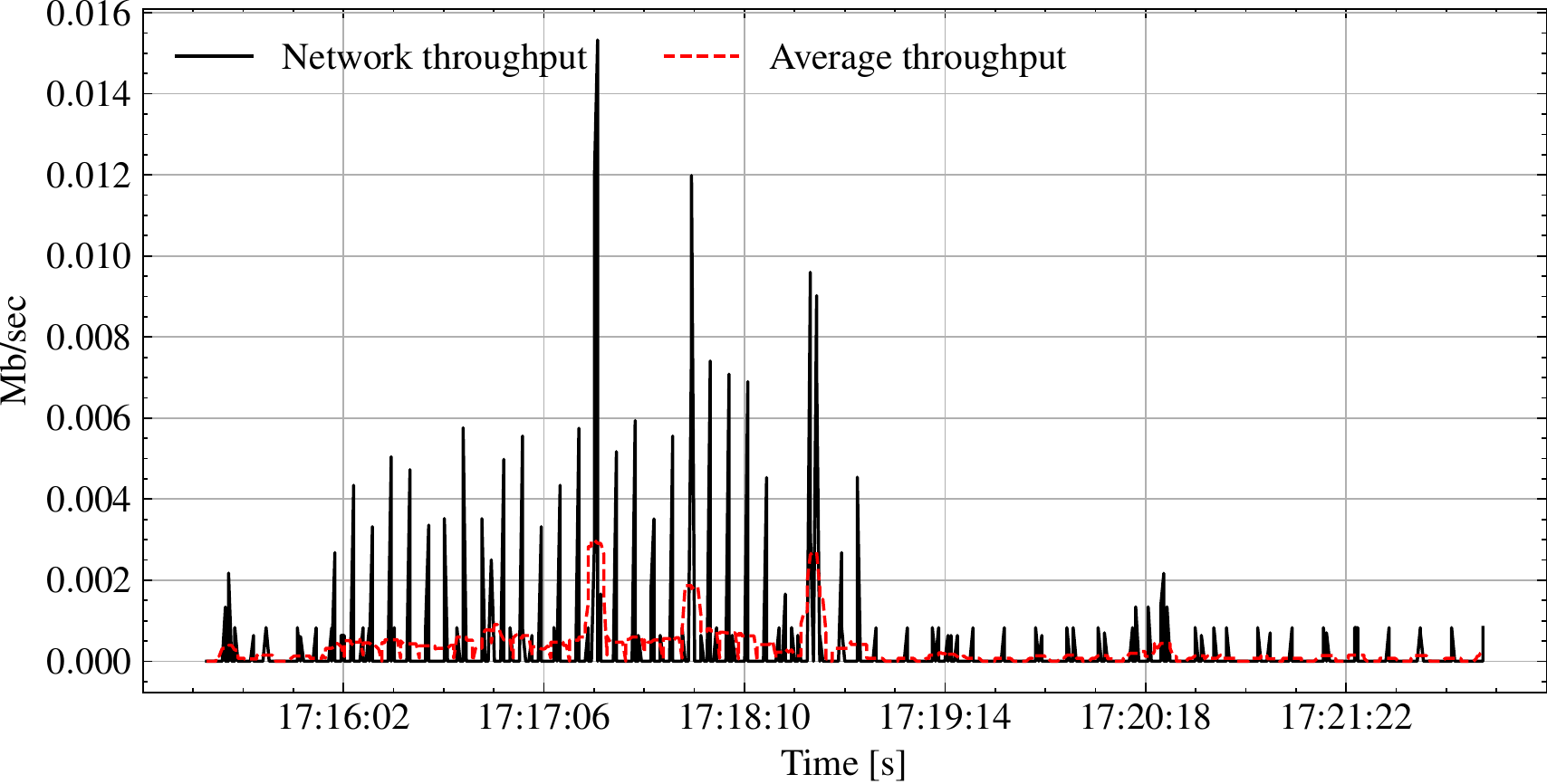}
        \caption{Message-level network throughput measured at the \gls{arx} node.}
        \label{fig:network-log-message}
    \end{subfigure}
    \caption{Network throughput for a) signal and b) message level relay. Minimum requirement is a) \SI{28}{Mb/s} network throughput, b) \SI{15}{KB/s} unless replaying is interrupted.}
    \label{fig:throughputs}
\end{figure}

Because of the large amount of data transferred between the attacker's nodes, network quality is important to guarantee a successful attack. 
From the performed experiment we observe that a minimum speed of \SI{27.5}{\mega \bit/\second} is required for the signal level replay to succeed. This is slightly lower than the calculated \SI{32}{\mega \bit/\second} for a complex signal at \SI{1}{Msps} sampling rate and 16 bit quantization. 
If this requirement is not met, we observe that the replay radio buffer is not able to receive enough data from the sampler to produce valid GNSS signals (\cref{fig:network-log}), causing a temporary loss of lock in the victim receiver.

During the tests we observed certain locations with degraded LTE coverage where the throughput of the cellular connection was sub-optimal. This event is visible at $t=11:14:15$ in \cref{fig:meacon-phases}. If the replay is interrupted for prolonged periods (in the order of the jamming period), it is possible that the attacker looses control over the victim receiver. On the other hand, as the network interruptions are temporary, the intermittent replay action results in jamming of the victim until a stable throughput is not restored. If the throughput is not impacted drastically, the non-continuous sample stream causes enough signal interference with the legitimate signals, that the victim does not regain lock on the latter. Possibly, such issue could be solved by focusing on rate constant transmission instead of minimal latency by sample buffering at the \gls{atx} node. Further investigation in this direction is ongoing.

u

\textbf{Message level relaying/replaying} experimentation brought up two major problems: for one, the reliable tracking and extracting of \gls{gnss} signals and parameters with the \gls{sdr} and the software receiver at the \gls{arx} node, and secondly, the generation of realistic spoofing signals at the \gls{atx} node.

While tracking the legitimate \gls{gnss} signals to extract the required signal parameters with \gnsssdr{}, we observed a high level of instability, presumably caused by the \gls{sdr} clock. We found that the \textit{LimeSDR} experienced significantly less loss of locks on legitimate satellites, compared to the BladeRF 2.0 \gls{sdr}, even when configured with an externally disciplined clock. 
Constant loss of lock reported by \gnsssdr{} when working with the BladeRF as front-end, can, among others, indicate problems in frequency stability due to oscillator imprecision. This effect is further amplified due to the cold temperature during our outdoor experiments. We therefore relied on the LimeSDR, featuring a more robust clock circuitry for our experiments.
We are currently investigating possible causes by evaluating the radio's performance in sampling rate and frequency stability. 

At the \gls{atx} node, we experienced a significant degradation in signal quality, when attempting to generate signals in real-time from a moving \gls{arx} node. 
Investigation into the signal regeneration component showed that \gls{pvt} X,Y,Z offsets between simulation steps greater than \SI{20}{m}, prevent demodulation in the victim receiver \footnote{At the current state, it is unclear whether this is caused by the inability of the victim receiver to track the resulting rapidly varying constellation, or if \gls{atx} signals are rejected based on internal spoofing countermeasures.}. 
By tuning \gnsssdr{} parameters (especially by employing carrier smoothing in the Observables block and using a more precise PVT implementation for mobile receivers), we were able to reduce the position offsets below \SI{1}{m}. This proved to be sufficient for the victim receiver to lock onto \gls{atx} signals, but introduced additional computation load for the \gls{arx} node.

\begin{figure}[htb]
    \centering
    \begin{subfigure}{0.48\textwidth}
        \includegraphics[height=7cm]{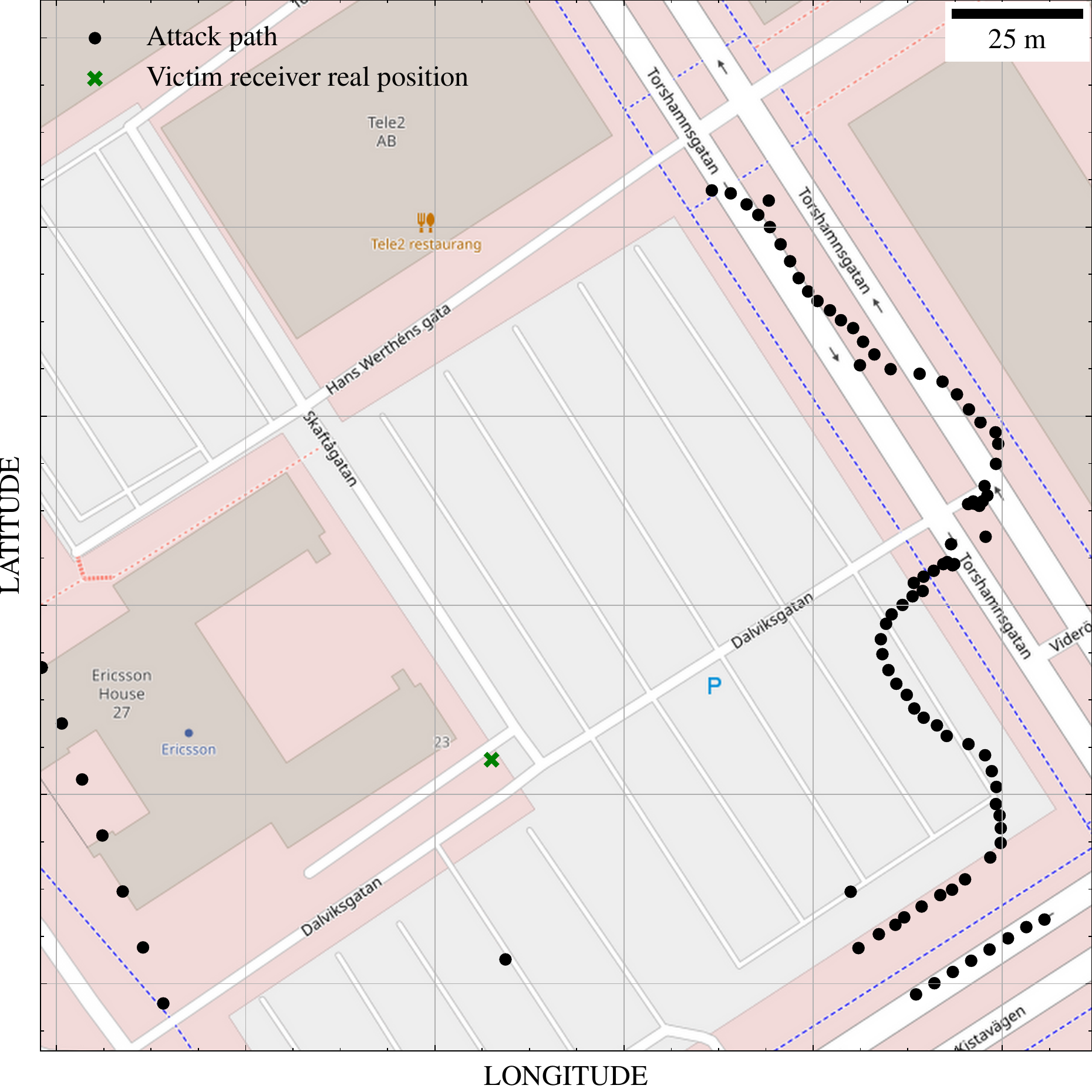}
        \caption{Attacker-induced position compared to actual victim position.}
        \label{fig:success-ref-message}
    \end{subfigure}
    \begin{subfigure}{0.48\textwidth}
        \includegraphics[height=7cm]{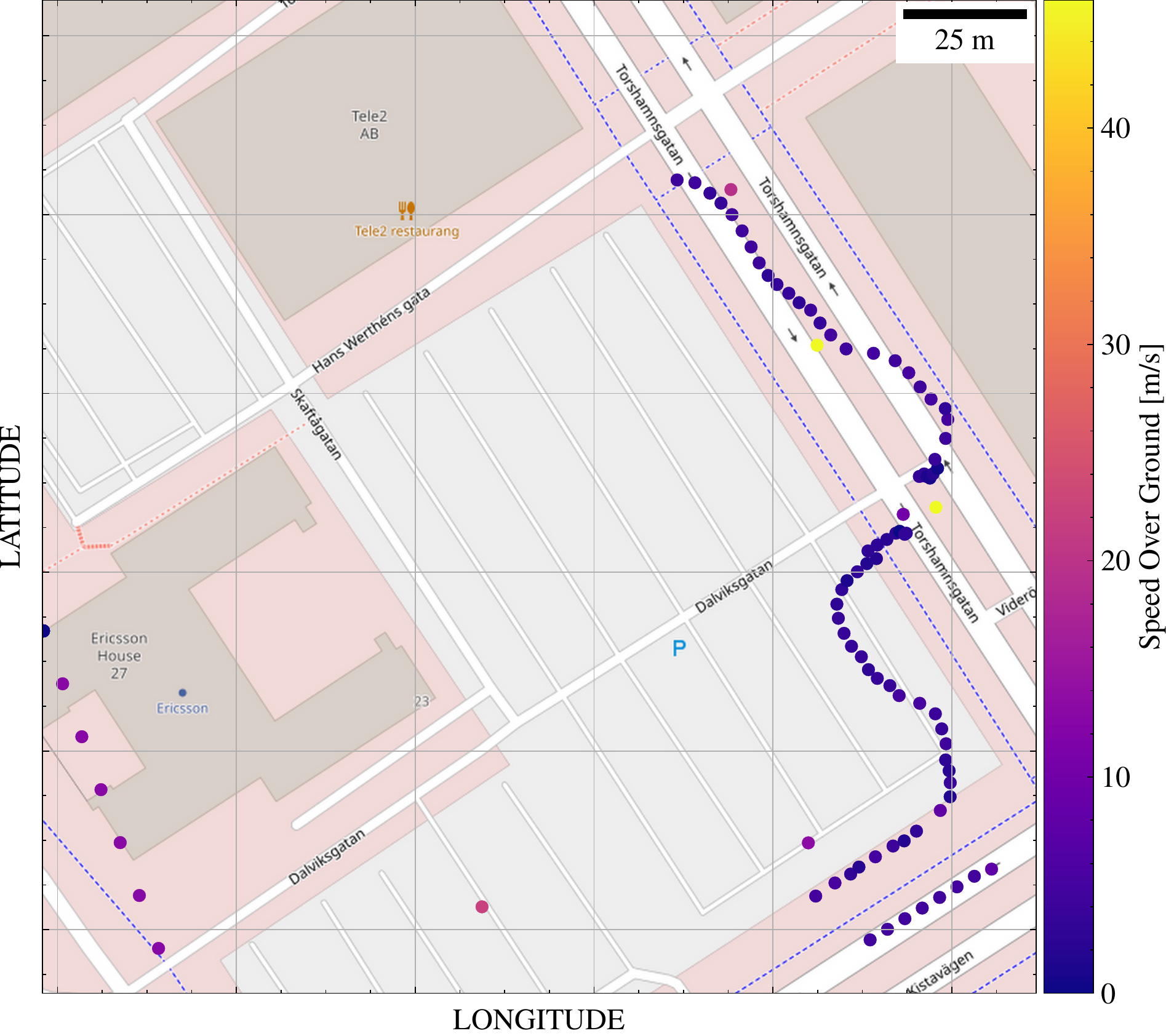}
        \caption{Attacker-induced velocity.}
        \label{fig:success-velocity-message}
    \end{subfigure}
    \caption{Successful message level relay/replay: the \gls{pvt} solution at the victim follows the \gls{arx} movement.}
    \label{fig:success-meacon-message}
\end{figure}

\begin{figure}[htb]
  \centering
  \begin{subfigure}[t]{0.48\textwidth}
    \includegraphics[width=\linewidth]{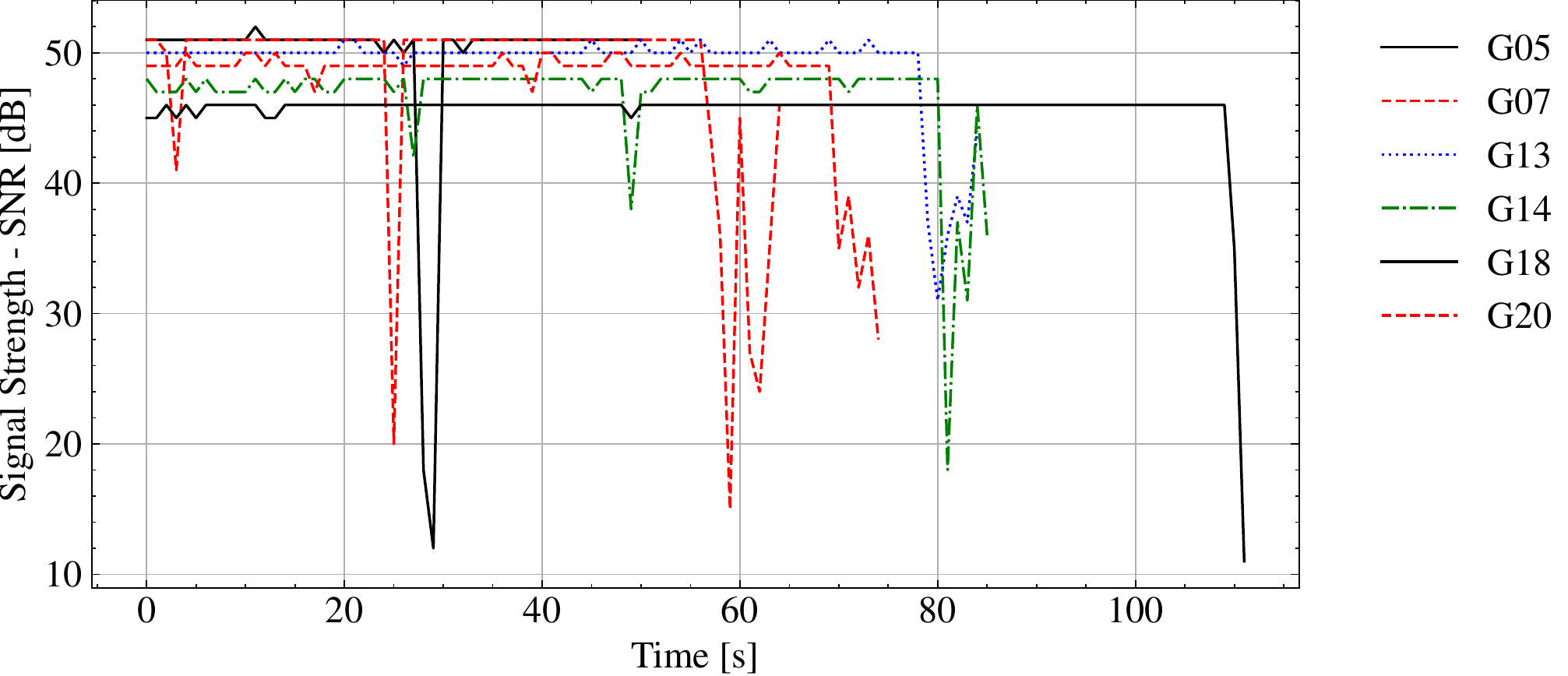}
    \caption{Message level replay/relay: SNR of the reconstructed messages at the victim receiver.}
    \label{fig:snr-victim-message}
  \end{subfigure}
  \hfill%
  \begin{subfigure}[t]{0.48\textwidth}
    \includegraphics[width=\linewidth]{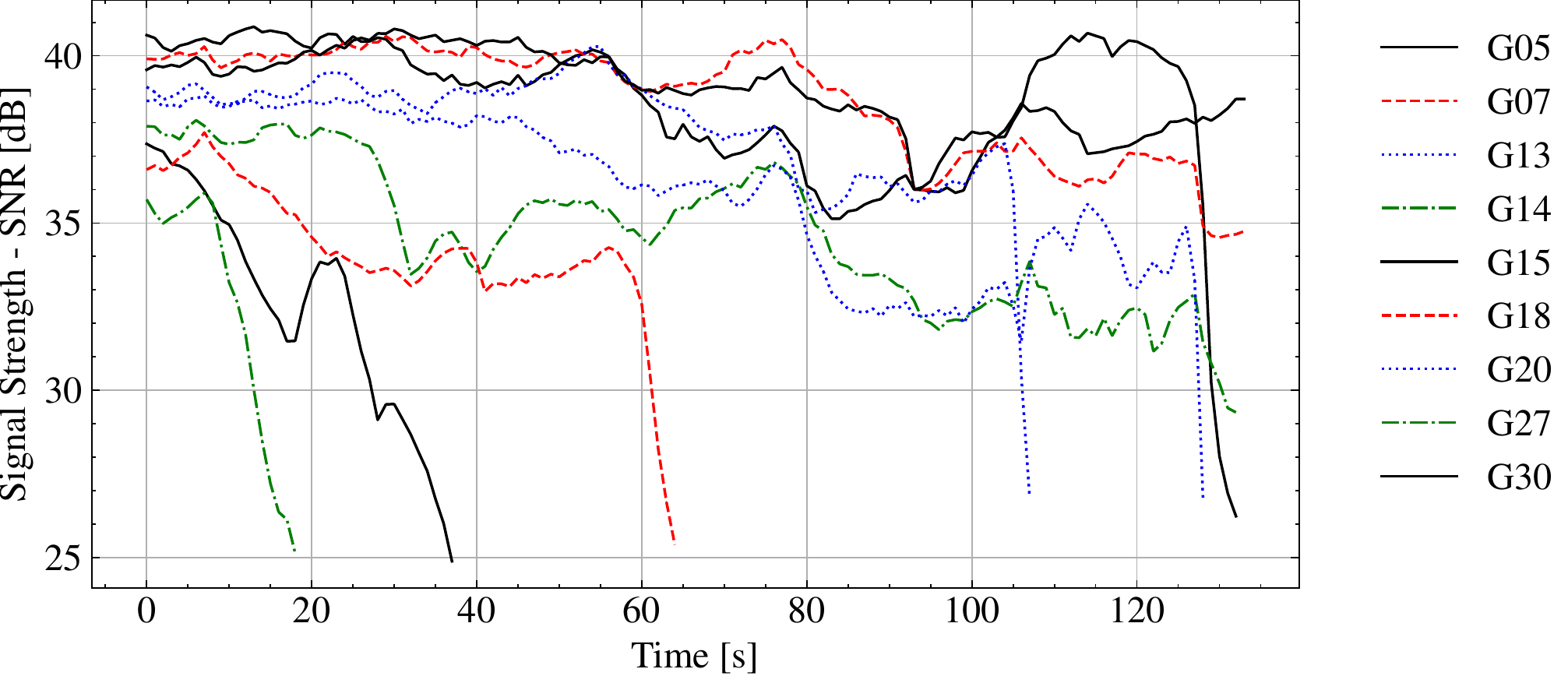}
    \caption{SNR at the \gls{arx} node. }
    \label{fig:snr-ref-sampler-message}
  \end{subfigure}
  \caption{SNR raw measurements at the victim and \gls{arx} node.}
   \label{fig:snr-attack-vs-ref-message}
\end{figure}

Our message level replayer 'captured' the victim \gls{gnss} receiver as depicted in \cref{fig:success-meacon-message}. The victim position successfully follows the position of the \gls{arx} node, although with lower accuracy than in the signal level replay. \cref{fig:success-velocity-message} shows that the speed of the victim receiver is in principle also determined by the \gls{arx} speed, with accuracy limited by the precision in signal reconstruction. 
The signal regeneration fails in places where the \gls{pvt} solution error at the \gls{arx} node increases. This is one possible reason why the depicted attack in \cref{fig:success-meacon-message} fails in the middle of the \gls{arx} path. Other plausible causes are degraded signal reception due to environmental shadowing, the \gls{arx} node speed, as well as mechanical effects due to a speed bump in the exact location where the attack fails. 
In our ongoing work, we investigate improvements on accuracy, e.g., more precise signal feature extraction, or by incorporating signal parameters, such as carrier-to-noise ratios into the signal regeneration. 
The \gls{snr} of the relayed/replayed signals do not match the one measured at the \gls{arx} node, as can be seen in \cref{fig:snr-victim-message} (\gls{atx} \gls{snr}) and \cref{fig:snr-ref-sampler-message} (legitimate signal \glspl{snr}). As only a subset of the satellites, as selected by the adversary, are relayed/replayed, characteristic satellites which disappear from view during the recording, are not visible in the replayed signals. 
Termporary \gls{snr} degradation in \cref{fig:success-ref-message} are due to \gls{pvt} loss at the \gls{arx} node or to outages in the network.

Compared to signal level replay, message level replay introduces processing delays in the \gls{arx} and \gls{atx} nodes. The \gls{arx} node processes individual satellites before distributing data to the \gls{atx} nodes, which rely on the \gls{pvt} solution calculated at the \gls{arx} node to generate spoofing signals. Without custom optimization, the \gnsssdr{} implementation for processing navigation messages and telemetry introduces a \SI{600}{\milli \second} delay (a full navigation sub-frame). Notably, further optimization in \gnsssdr{} can optimize and reduce this latency, e.g., streaming the navigation message bit or word wise. This would, however, require a higher degree of synchronization between the colluding adversaries, as well as better error correction. Further improvements on the replay latency are out of the scope of this work and are left for future work. 
The only resulting difference caused by a longer processing delay from an attackers point of view, is the length of the required jamming phase until the receiver's time uncertainty allows acceptance of the spoofed signals \cite{teunissenSpringerHandbookGlobal2017}.

We highlight the bandwidth reduction achieved by the message level replay :
\cref{fig:network-log-message} shows how the required throughput for the \gls{arx}-to-\gls{atx} peaks at \SI{15}{KB/s}. This brings it well within reach of most cellular plans, thereby increasing the flexibility of distributed relay/replay in highly mobile scenarios, possibly with both the attacker and the victim moving.

\section{Conclusion}
\label{sec:conclusion}
We demonstrated the capabilities of a network-based \gls{gnss} relayer/replayer, that operates at signal level and at message level. We developed an experimental setup that can be used for future research on relay/replay spoofing countermeasures.  Its modular design allows different attacker/victim configurations.
By basing the attacker on \gls{cots} devices, we facilitate replication by researchers to investigate detection schemes with legitimate and upcoming authenticated GNSS signals. An attack demonstration on \gls{cots} hardware furthermore stresses the present day risk for spoofing incidents.

Our results show that signal level replay over a mobile network is feasible, provided that the wireless \gls{arx}-\gls{atx} connection has sufficient throughput. With high-end consumer grade cellular connectivity, as the data plan used in our experiments, this is a realistic assumption, especially with the further deployment of fast cellular technologies. 
Our successful message level relay/replay attack design, by overcoming bandwidth limitations,
demonstrate that relaying/replaying attacks, especially by colluding networked attackers, pose a versatile threat to the security of \glspl{gnss}. 

In addition to addressing open questions discussed in \cref{sec:evaluation}, we plan to include recently introduced authenticated Galileo \gls{os-nma} signals into our experiments.
Our test-bench has been designed with the integration of authenticated signals in mind, requiring only minor adaptation in the \gls{arx} node, and custom re-generation units for each targeted \gls{gnss} band. Another avenue for future work is the inclusion of more advanced attack types. Message level relaying/replaying allows to attack multiple \gls{gnss} bands simultaneously, and allows integrating more advanced attacks, e.g., introducing selective delays to specific satellite streams.

\section*{acknowledgements}
This work was supported in part by the SSF SURPRISE cybersecurity project and the Security Link strategic research center.

\bibliographystyle{IEEEtran}
\bibliography{biblio}

\end{document}